\documentclass[%
aip,jcp,amsmath,amssymb,floatfix,reprint,citeautoscript
]{revtex4-1}

\usepackage{graphicx}%
\usepackage{dcolumn}%
\usepackage{bm}%

\usepackage[utf8]{inputenc}
\usepackage[T1]{fontenc}
\usepackage{mathptmx}
\usepackage{etoolbox}
\usepackage[colorlinks,allcolors=black,citecolor=blue,urlcolor=blue]{hyperref}

\usepackage{comment}

\makeatletter
\def\@email#1#2{%
 \endgroup
 \patchcmd{\titleblock@produce}
  {\frontmatter@RRAPformat}
  {\frontmatter@RRAPformat{\produce@RRAP{*#1\href{mailto:#2}{#2}}}\frontmatter@RRAPformat}
  {}{}
}%
\makeatother

\graphicspath{{figures/}}

\newcommand{\phm}{\ensuremath{\textit{p}\rm{H_2} }}
\newcommand{\hhm}{\ensuremath{\rm{H_2} }}
\newcommand{\wat}{H$\rm _2$O}
\newcommand{\pwat}{H$\rm _3$O$^+$}
\newcommand{\ph}{\textit{p}H$\rm _2$}
\newcommand{\h}{H$\rm _2$}

\begin{document}

\preprint{AIP/123-QED}

\title[
Interaction Potentials for \textit{para}-Hydrogen 
with 
Flexible Molecules
]{
Neural Network Interaction 
Potentials for \textit{para}-Hydrogen
with Flexible~Molecules
}

\author{Laura Dur\'an Caballero}
\email{laura.duran-caballero@rub.de}

\author{Christoph Schran}
\altaffiliation[Present address: ]{Yusuf Hamied Department of Chemistry, University of Cambridge, Cambridge, CB2 1EW, UK}

\author{Fabien Brieuc}
\altaffiliation[Present address: ]{Laboratoire Mati\`ere en Conditions Extr\^emes,
Universit\'e Paris-Saclay, CEA, DAM, DIF, 91297 Arpajon, France}

\author{Dominik Marx}
\affiliation{Lehrstuhl f\"ur Theoretische Chemie, Ruhr-Universit\"at Bochum, 44780 Bochum, Germany}

\date{\today}%
\begin{abstract}
The study of molecular impurities in \textit{para}-hydrogen (\ph{}) clusters
is key to push forward our
understanding of intra- and intermolecular interactions
including their impact on the superfluid response of this bosonic quantum solvent.
This includes tagging
with only one or 
very few \ph{}, 
the microsolvation regime 
for intermediate particle numbers, 
and
matrix isolation
with many solvent molecules.
However, the fundamental coupling between the bosonic
\ph{} environment and the \mbox{(ro-)vibrational} motion
of molecular impurities remains poorly understood.
Quantum simulations can in principle provide the necessary atomistic insight,
but they require very accurate descriptions of the involved interactions.
Here, we present a data-driven approach for the generation
of
\mbox{\textit{impurity}$\cdots$\ph{}} interaction
potentials based on machine learning techniques
which retain the 
full flexibility of the dopant species. 
We
employ the well-established adiabatic hindered rotor (AHR) averaging technique
to include
the impact of the nuclear spin statistics 
on the symmetry-allowed rotational quantum numbers
of \ph{}.
Embedding this averaging procedure within the 
high-dimensional 
neural network potential (NNP) framework
enables the generation of highly-accurate AHR-averaged NNPs
at coupled cluster accuracy,
namely \mbox{CCSD(T$^*$)-F12a/aVTZcp},
in an 
automated 
manner.
We apply this methodology to 
the water and protonated water molecules,
as representative cases for quasi-rigid and highly-flexible molecules
respectively, 
and obtain 
AHR-averaged NNPs
that reliably describe the corresponding \mbox{\wat{}$\cdots$\ph{}} and 
\mbox{\pwat{}$\cdots$\ph{}} interactions.
Using path integral simulations
we show
for the 
hydronium cation, \pwat{}, 
that umbrella-like tunneling inversion 
has a strong impact on the first and second
\ph{} microsolvation shells.
The automated and data-driven nature of our protocol opens 
the door to the study of bosonic \ph{} quantum
solvation for a wide range of 
embedded impurities. 
\end{abstract}

\maketitle

\section{Introduction}
It 
has been
known for a long time that $^4$He is superfluid at
temperatures below 2.17\,K~\cite{
kapitza1938viscosity, 
allen1938flow}. 
In the search for other substances that could be superfluid, \textit{para}-hydrogen (\ph{}) 
--- a zero-spin boson
characterized by even values of the rotational quantum number~---
was considered very early
to be a suitable candidate~\cite{
ginzburg1972can},
yet experimental 
evidence of superfluid behavior of 
OCS-doped
\ph{} clusters 
embedded in helium nanodroplets
was 
only reported~\cite{grebenev2000evidence} in 2000 
although it has been predicted computationally for pure \ph{} clusters roughly ten years earlier~\cite{
sindzingre1991superfluidity}. 
Note that a different interpretation of the experimental signatures
has been advanced in 2019 based on
computational evidence for the valence-isoelectronic CO$_2$ dopant~---
culminating in the conclusion that
CO$_2$-doped \ph{} clusters embedded within $^4$He clusters feature
minimal superfluid
response but instead phase-separate and form a nonsuperfluid
\mbox{CO$_2$(\ph{})$_N$} core therein~\cite{
li2019supression}. 
Independently from this 
discussion,
the bosonic nature of \ph{} together with its low mass make it a
perfect candidate to undergo a superfluid transition. 
While bulk \ph{} remains solid even at ultra-low temperatures
due to the relatively strong intermolecular interactions~\cite{silvera1980solid},
pure
\ph{} clusters of specific size
were
predicted to 
feature manifestations of superfluidity 
by various quantum 
simulations~\cite{
sindzingre1991superfluidity,
scharf1992path,
mezzacapo2006superfluidity,
mezzacapo2007structure,
khairallah2007interplay,
mezzacapo2008local,
cuervo2006path,
li2010molecular,
zeng2014microscopic,
schmidt2015raman,
schmidt2022ground}.
However, experimental verification of these effects can only be indirectly achieved
and, additionally, might be challenging to correctly interpret.
But given
the routine use of \ph{} in 
both,
matrix isolation spectroscopy~\cite{
bondybey1996new,
momose2005chemical}
and vibrational photodissociation spectroscopy using messenger tagging~\cite{
okumura1986infrared,
chakrabarty2013novel,
wolk2014cryogenic}
it is necessary to 
advance our
microscopic understanding 
of the intermolecular interactions between \ph{} and molecular impurities 
as well as the possible coupling to superfluidity 
using accurate and 
predictive calculations.

In order to bring light into these questions, quantum simulation techniques 
are crucial to provide atomistic insights
as has been demonstrated since decades. 
Already in 1991,
pioneering
Path Integral Monte Carlo (PIMC) simulations of small 
\ph{} clusters predicted superfluid behavior for 13 and 18 molecules below 2~K~\cite{
sindzingre1991superfluidity}.
Although several studies on \ph{}~\cite{
sindzingre1991superfluidity,
scharf1992path,
mezzacapo2006superfluidity,
mezzacapo2007structure,
khairallah2007interplay,
mezzacapo2008local,
cuervo2006path,
li2010molecular,
zeng2014microscopic,
schmidt2015raman,
schmidt2022ground},
\textit{ortho}-deuterium~\cite{
scharf1999isotope,
mezzacapo2007structure,
cuervo2008solid}, 
and mixed~\cite{
gordillo1999binary, 
cazorla2008two,
cuervo2009weakly} 
clusters have been performed since then,
theoretical studies regarding molecular impurities in \ph{}
have mainly been limited to 
small neutral molecules including,
among others, 
$\rm OCS$~\cite{
kwon2002nanoscale}, 
$\rm CO_2$~\cite{
lara-castells2011collective,
li2011theoretical,
li2019supression}, 
$\rm CO$~\cite{
raston2012persistent},
$\rm CH_4$~\cite{
mak2005superfluidity},
$\rm SO_2$~\cite{
zeng2013probing}
and $\rm H_2O$~\cite{
zeng2012simulating}
as reviewed in Ref.~\citenum{zeng2014microscopic}. 
Importantly, all these studies have been performed while freezing
the nuclear skeleton of the molecular impurities, thus treating
them as (quantum) rigid rotors embedded in \ph{} clusters 
while neglecting the 
vibrational degrees of freedom~\cite{zeng2016moribs}.
Simulating non-atomic impurities in bosonic environments as rigid bodies 
is expected to be 
an almost perfect 
approximation for the plethora 
of quasi-rigid molecules such as OCS, CO$_2$, CH$_4$ or H$_2$O, 
but this assumption might introduce severe artifacts
when investigating non-covalently bound molecular complexes~\cite{
walewski2014reactive, 
walewski2014solvation}
and even more so for floppy molecules~\cite{
witt2013microsolvation,
uhl2018helium,
uhl2019quantum}
all belonging to the family of non-rigid systems.

To date only very few simulation techniques exist
to go beyond the rigid rotor treatment of molecular impurities
in superfluid quantum solvents at ultra-low but finite temperatures.
The hybrid 
path integral molecular dynamics/bosonic path integral Monte Carlo (HPIMD/MC)
technique~\cite{
walewski2014reactive, 
walewski2014solvation} 
developed in our group is one of the few examples,
as recently reviewed in Ref.~\citenum{brieuc2020converged}.
The combination of both approaches makes it possible to run converged 
path integral simulations at ultra-low temperatures
including the bosonic character of the quantum solvent
(via bosonic exchange 
PIMC sampling) 
while keeping the flexibility of the dopant molecule
(via PIMD propagation). 
This methodology has so far been used to study reactive \mbox{HCl$\cdots$water}
clusters in superfluid helium nanodroplets~\cite{
walewski2014reactive,
walewski2014solvation}
as well as
protonated methane 
\cite{
uhl2018helium, 
uhl2019quantum} 
and protonated water clusters~\cite{
schran2018high}
microsolvated by $^4$He atoms.

Unfortunately, the application of this finite-temperature methodology to
\ph{} is not straightforward, as accounting for bosonic exchange
requires treating the quantum solvent as 
identical 
point-like particles
using present-day simulation techniques to sample symmetrized path integrals
in order to establish Bose-Einstein statistics. 
Traditionally, an isotropic spherical average over the orientation of 
the two-site \h{} molecule 
has
been performed~\cite{
kwon2002nanoscale,
paesani2003ocs, 
paesani2005ocs, 
kwon2005superfluid,
moroni2005small,
piccarreta2006structuring}
to generate a spherical single-site interaction potential for \ph{}. 
However,
it has been shown later 
that a non-negligible error is made 
when using this approach to describe the interaction potential between
\ph{} and molecular impurities, especially for strongly
interacting and anisotropic systems, such as 
the
water molecule~\cite{
zeng2011adiabatic}. 
The so-called
`adiabatic hindered rotor' (AHR) approximation~\cite{
li2010adiabatic,
zeng2011adiabatic} 
has been devised to
overcome this problem by 
effectively including the anisotropy
of the interaction potential into the averaging procedure.
This allows one to treat \ph{} in bosonic quantum simulations 
as if it were a spherical point particle.

Although 
reliable and even
highly accurate 
pair potentials for the description of the 
\mbox{\ph{}$\cdots$\ph{}}
interaction are available in the literature,
such as the semiempirical Silvera-Goldman pair potential~\cite{
silvera1978isotropic}
and electronic-structure-based two- and three-body potentials~\cite{
patkowski2008potential,
ibrahim2022three}, 
it 
remains
difficult to obtain accurate interaction potentials between 
dopant molecules and \ph{}
despite much progress~\cite{
li2010adiabatic, %
zeng2011adiabatic, %
zeng2013probing, %
wang2013new, %
zhang2018analytic}. %
The generation of such potentials using physically motivated
functional forms is usually very challenging and a time-consuming task
in part because the functions used are rarely transferable 
from one system to another.
Machine learning techniques have been shown to be a promising alternative~\cite{
behler2021four,
manzhos2021neural}
for the current task 
for which most steps of the development can be easily automated~\cite{
schran2018high},
and transferred to a new system
while using 
basically converged coupled cluster accuracy
at the level of the reference calculations.

In this work, 
we introduce a general methodology to generate
highly accurate AHR-averaged
potentials 
to describe
the interaction between a molecular impurity and \ph{}  
in the framework of 
high-dimensional 
neural network potentials (NNPs)~\cite{
behler2007generalized,
behler2017first,
behler2021four}.
Our data-driven approach consists of a 3-step process: 
First an all-atom NNP is generated for the 
interaction potential
between \h{} and the dopant molecule,
where \h{} is treated as a 
diatomic
molecule.
Next the AHR-average is performed using this all-atom NNP
in an automated procedure
to generate the effective 
interactions of point-like \ph{}
with the impurity. 
Finally, a NNP representation of the 
\mbox{\textit{dopant}$\cdots$\ph{}}
interaction potential is obtained based on 
this effective single-site treatment
of \ph{}.
We demonstrate the capabilities of this data-driven procedure
for the interaction potential of \ph{} with water, for which
existing AHR potentials are available for reference~\cite{zeng2011adiabatic}
where H$_2$O has been described as a rigid body.
We also apply it 
to
the
distinctly non-rigid and 
strongly interacting
hydronium cation (\pwat{}),
where the inclusion of the full flexibility of the 
molecular dopant  
is crucial
as will be demonstrated.
Importantly, our AHR-averaged NNP interaction potential allows one to include the 
(ro-vibrational) flexibility of these two impurities at the level
of bosonic HPIMD/MC quantum simulations. 
Overall, this 
work opens up the possibility for the computational study of
utmost floppy and even reactive
impurities under bosonic \ph{} solvation
to shed light on the coupling of superfluid behavior
and intermolecular interactions
as disclosed previously for the highly fluxional protonated methane molecule (CH$_5^+$)
interacting with bosonic helium~\cite{uhl2018helium,uhl2019quantum}.

\section{Methodology}
A key aspect of most computational approaches for 
investigating
molecular impurities in bosonic quantum solvents 
is to describe 
the full system as the sum of \mbox{\textit{solute}$\cdots$\textit{solvent}}, 
\mbox{\textit{solvent}$\cdots$\textit{solvent}}
and possibly intra-solute interactions 
which allows the simulations to be computationally efficient. 
Performing such simulations using \ph{} as the quantum solvent
requires the treatment of the 
\ph{} molecules as spherical point-like particles to practically include bosonic exchange. 
In order to generate an interaction potential between the molecular impurity
and the point-like \ph{}, it is necessary to 
use a suitable averaging procedure over the orientation of the \ph{} molecule
as already alluded to in the introduction. 
It has been shown that the AHR approximation~\cite{
li2010adiabatic, 
zeng2011adiabatic,
wang2013new,
zhang2018analytic} 
can be used to perform such an 
anisotropic
average 
including the 
allowed rotational levels 
to obtain a reduced-dimension interaction potential
where \ph{} is an effective spherical point particle.
Here, this well-established approach will be
combined with machine learning methodology in order to 
produce highly-accurate 
\mbox{\textit{solute}$\cdots$\textit{solvent}}
interaction potentials 
of fully flexible and possibly reactive molecular dopants with a point-like \ph{} molecule.

\begin{figure*}[]
\centering
\includegraphics[width=\textwidth]{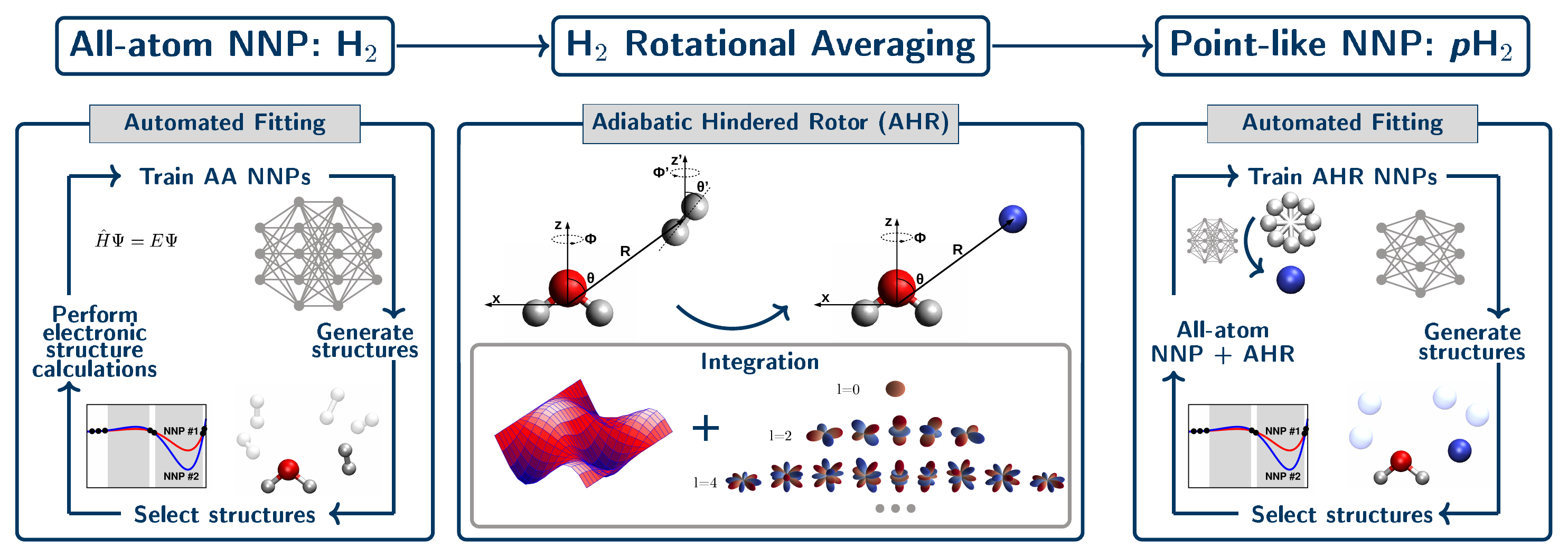}
\caption{%
  Schematic representation of the data-driven 3-step development process for
  generating all-atom \h{} and single-site AHR-averaged \ph{} neural network interaction potentials.
  The left panel shows the generation of the 
interim 
all-atom 
interaction NNP 
  via an 
automated 
active learning procedure
where the diatomic nature of the \h{} molecule is explicitly considered. 
  Then, that final all-atom NNP is used to perform the 
  adiabatic hindered rotor~(AHR) averaging (middle panel).
  In this step we reduce the dimensionality of the \ph{} molecule
  from all-atom to a point-like particle,
  while imposing the required nuclear spin statistics for \ph{} 
  by only including the even rotational energy levels in the orientational average. 
  Finally, the AHR-averaged 
  NNP 
  is generated, where \ph{} is a
  point-like
 and thus effective spherical 
particle (right panel).
} 
\label{fig: overview}
\end{figure*}

Our data-driven process consists of three steps.
First, an all-atom NNP for the
interaction potential
between \h{} and the impurity of interest is generated,
where the rotational degrees of freedom of the \h{}
molecule are treated explicitly
while its bond length is fixed at its
vibrational average in the ground state (0.7668~\AA)~\cite{diep2000accurate}.
The all-atom NNP is fitted to highly-accurate coupled cluster
electronic structure calculations.
Secondly, this all-atom NNP
is used to most efficiently perform the AHR averaging using that interim interaction potential 
in order to reduce the \ph{} molecule to 
an effective spherical point particle and thus to a single interaction site. 
Finally, a 
single-site
NNP is generated for the point-like particle representation of the 
\ph{}
interaction potential, in which the obtained AHR-averaged 
NNP
energies are used as reference data
to fit the final AHR-averaged (single-site) NNP. 
A schematic representation of this 3-step process can be found
in Figure~\ref{fig: overview}.

Our machine learning approach builds on the concepts of
high-dimensional neural network potentials (NNPs) combined with
atom-centered symmetry functions~\cite{
behler2011atom} 
as developed by Behler and coworkers~\cite{
behler2007generalized, 
behler2011atom, 
behler2017first,
behler2021four}.
This methodology first transforms the atomic coordinates using
a set of symmetry functions~\cite{
behler2011atom} 
into suitable input for atomic neural networks.
These atomic neural networks output atomic energy contributions such that the total
high-dimensional NNP interaction energy is the sum of the atomic contributions 
obtained by the atomic NNs.

The generation of such interaction NNPs 
can be automated
and tuned to efficiently select the configurations for 
optimal training of the NNP so that the number of expensive
reference electronic structure calculations is minimized,
as introduced and described in detail in 
Ref.~\citenum{
schran2018high}
for the case of helium interacting with protonated water clusters. 
This 
active-learning
procedure is based on two strategies in order to efficiently
select the structures that are added to the reference dataset,
by using the flexibility of neural networks to our advantage.
On one side, this flexibility can be used to detect underrepresented
regions of the configuration space. 
This is achieved by comparing two differently trained NNPs
and identifying the configurations that feature the largest disagreement.
On the other side, extrapolation of the NNP can be easily detected so that
configurations outside of the sampled configuration space
can be added to the training set, thus expanding its boundaries.
The all-atom interaction NNP is generated using our
automated 
procedure~\cite{
schran2018high}, 
but adapted from the single-site He~atom to the two-site \h{} molecule 
in order to include the additional
degrees of freedom coming from the orientation 
of the diatomic molecule around its center of mass. 
Furthermore, we now make use of the advantages of committee approaches~\cite{
schran2020committee}
to achieve an improved accuracy of our new interaction NNPs.
Once the dataset is generated, different NNPs are fitted to form a committee NNP, 
where the prediction of the model is given by the average over the committee members.
This approach is widely used in the neural network community
as it improves the accuracy of the models 
at not much additional computational cost and also since it
provides 
a
generalization 
of
error estimates.
For more 
background and
details on this approach we refer the interested reader to Ref.~\citenum{
schran2020committee}.

Once the all-atom NNP for the interaction 
of \h{} with a specific molecular impurity 
is generated, the AHR approximation~\cite{
li2010adiabatic, 
zeng2011adiabatic,
wang2013new,
zhang2018analytic} 
is used to reduce the dimensionality of the \ph{} molecule
while incorporating the required nuclear spin statistics by considering
only the even rotational levels in the rotational average.
The AHR approximation is based on the assumption that the \h{} rotation is 
much faster than the motion of the other degrees of freedom, and can 
therefore be 
adiabatically
separated and averaged out by considering only the 
allowed rotational states, i.e. symmetric w.r.t permutations of the two protons for \ph{}. 
The adiabatic Schr\"odinger equation
$$ \left( \hat{T}_{\hhm} + \frac{\hbar^2}{2\mu R^2} \hat{j}^2_{\hhm} 
+ V(R,\theta,\phi,\theta^\prime,\phi^\prime) \right) \psi 
(\theta^\prime, \phi^\prime; R, \theta, \phi) $$
\begin{equation}
= V^{\rm{AHR}} (R, \theta, \phi)\ \psi (\theta^\prime, \phi^\prime; 
R, \theta, \phi)\ ,
\label{eq: adiabatic schrodinger equation}
\end{equation}
is solved, where the lowest eigenvalue is the AHR-averaged potential 
$V^{\rm{AHR}} (R, \theta, \phi)$ at a given position of the \ph{} molecule, 
$ V(R,\theta,\phi,\theta^\prime,\phi^\prime)$ 
is the all-atom interaction potential (obtained here from the all-atom NNP), 
$\hat{T}_{\hhm}$ is the kinetic energy operator of \h{},
$\hat{j}^2_{\hhm}$ is the angular momentum operator of \h{} 
and $\mu$ is the reduced mass 
of the \mbox{\textit{impurity}$\cdots$\ph{}} pair; 
see the middle panel of Figure~\ref{fig: overview} for the definition of the required coordinates
$(R,\theta,\phi,\theta^\prime,\phi^\prime)$. 
To solve this equation, spherical harmonics are used as basis functions to 
represent the
rotational
eigenstates
\begin{equation}
\displaystyle \psi (\theta^\prime, \phi^\prime; R, \theta, \phi) 
= \sum_{l,m} C_l^m ( R, \theta, \phi ) Y_l^m (\theta^\prime, \phi^\prime)
\label{eq: lc spherical harmonics}
\end{equation}
but 
considering only even values of $l$
together with the corresponding set of $m$ quantum numbers in the summation 
as required for \ph{};
note that including only the spherically symmetric $l=0$ term in that sum
yields the isotropic spherical quantum average of the interaction potential.

The contribution coming from the 
\h{} 
kinetic energy and angular momentum operators
to the AHR-averaged potential are easily obtained.
On the contrary, the evaluation of the contribution coming from the potential energy term
is more complicated, as it requires the computation of the following matrix elements
$$ \langle Y_{l^\prime}^{m^\prime} | V(R,\theta,\phi,\theta^\prime,\phi^\prime) | Y_l^m \rangle = $$
\begin{equation}
\int_0^\pi d\theta^\prime  \sin \theta^\prime\  
\int_0^{2\pi} d \phi^\prime\ Y_{l^\prime}^{m^\prime} (\theta^\prime,\phi^\prime)\ V(R,\theta,\phi,\theta^\prime,\phi^\prime)\ Y_l^m (\theta^\prime,\phi^\prime)\ .
\label{eq: integration potential term}
\end{equation}
These two integrals are computed numerically using Gauss-Legendre and Gauss-Chebyshev 
quadratures for $\theta^\prime$ and $\phi^\prime$, respectively,
while using the all-atom NNP. 

When performing the AHR average, we consider the coordinate system 
employed in 
previous pioneering work~\cite{
zeng2011adiabatic}.
It is 
defined such that the origin of coordinates is at the center of mass of the impurity, 
with its 
principal rotational 
axis aligned with the $z$ axis.
Then, an atom not belonging to the rotational axis is placed on the $xz$ plane.
A second coordinate system is used for 
\h{},
denoted here with $^{\prime}$, which has the same orientation as 
the impurity's coordinate system but its origin at the center
of mass of the 
\h{} molecule.
Here, 
$R$, $\theta$ and $\phi$ define the position of the center of mass of the 
\ph{} with respect to the origin of coordinates of 
the impurity's coordinate system, whereas
$\theta^\prime$ and $\phi^\prime$ define the orientation of the \ph{} molecule 
with respect to its center of mass.
This definition of the coordinate system for water is shown 
in the central panel of Figure~\ref{fig: overview}.
For more details on the AHR-averaging procedure, we refer the interested
reader to Refs.~\citenum{
li2010adiabatic,
zeng2011adiabatic,
wang2013new,
zhang2018analytic}.

The 
third and
last step in our data-driven protocol is to train a second NNP for the system where
\ph{} is treated as a point-like particle.
This AHR-averaged NNP is generated using the same automated procedure
as in the generation of the all-atom NNP introduced above, 
with the difference that now the AHR-averaged interaction energy 
$V^{\rm{AHR}} (R, \theta, \phi)$ 
from the 
solution of the
adiabatic Schr\"odinger equation 
Eq.~(\ref{eq: adiabatic schrodinger equation}) 
is used 
to provide the reference data
instead of electronic structure reference calculations
to which the interim all-atom NNP has been fitted to provide 
$V(R,\theta,\phi,\theta^\prime,\phi^\prime)$.
Therefore, the AHR-averaged NNP is directly 
obtained from
the all-atom NNP
through the AHR-averaged interaction energies.

\section{Computational details}
\label{sec:compdet} 
The first step is to produce the all-atom NNP to describe the
interaction between the solute molecule and \h{}.
This all-atom NNP is obtained starting from the 
potential energy surface that describes non-rigid solute species
(which we represent here in terms of another highly accurate, 
intramolecular~NNP published earlier~\cite{
schran2020automated}) 
in order to account for the flexibility of the molecule. 
The dataset is obtained using the automated procedure~\cite{
schran2018high}
developed for He~atoms, 
which has been extended to work with the \h{} molecule, 
namely to consider its possible orientations around the solute molecule. 
The original procedure employed to select the position of the helium atoms 
is used to select the position of the center of mass of the \h{} molecule, 
and its orientation is then randomly generated centered on that point;
note that this procedure can be readily generalized from diatomics
to polyatomic 
solvents
such as CH$_4$.
It is important that all possible orientations of the \h{} molecule
are properly represented in the all-atom NNP, 
since the 
subsequent 
AHR procedure averages over them and 
thus needs to rely on exhaustive angular information in order to 
accurately 
compute 
the matrix elements given by Eq.~(\ref{eq: integration potential term})
using the all-atom NNP to represent $V(R,\theta,\phi,\theta^\prime,\phi^\prime)$.
While building the dataset, all the configurations 
where \h{} is 
further away than 12~\AA{} 
from the solute
are excluded, 
as well as those for which the \h{} center of mass is
closer than 2.05~\AA{} to the O atom or 1.25~\AA{} to the H atoms
of the water and protonated water molecules,
similarly to previous work for the NNP interaction potential
of protonated water clusters in helium~\cite{
schran2018high}.

The dataset for the \mbox{\wat{}$\cdots$\h{}} all-atom NNP
has been generated with 100 loops of the automated procedure
adding 100 structures in each loop. 
The structures added in each loop have been sampled out of a set of
200~000 randomly generated configurations, which is reduced to
a 100 by our active learning procedure.
These pruning steps are urgently needed in order to reduce the computational
effort of the demanding coupled cluster electronic structure
calculation to a minimum of configurations while exhaustively sampling 
the relevant configuration space to fit the NNP.
The generated dataset of structures is then 
randomly
divided into a 
training
set composed of 90~\% of the selected structures and
a test set containing the remaining 10~\%.
The fits of the NNP have been performed with the \texttt{RubNNet4MD} program~\cite{
rubnnet}
using an architecture of 3~hidden layers with 30~nodes each. 
A hyperbolic tangent activation function is used in all hidden layers
and a linear one for the output layer. 
The weights of the NN are optimized 
by the element-decoupled Kalman 
filter~\cite{blank1994adaptive,%
gastegger2015high-dimensional}.
A total of 10~different 
independent
models have been generated that way and the committee approach 
\cite{
schran2020committee}
has been applied to obtain the final all-atom NNP
as their 
average.
The same approach is used to generate the all-atom NNP
for \mbox{\pwat{}$\cdots$\h{}}, but only 60~loops of the automated
procedure 
were
performed resulting in 
a grand total of 
59~800 selected structures.
The all-atom and AHR-averaged final interaction NNPs
for \wat{} and \pwat{} interacting with \h{} and \ph{}, respectively,  
are provided as Supplementary Information 
to this paper.

The reference interaction energies have been computed using the 
coupled cluster method~\cite{
cizek1966correlation,
bartlett1989coupled,
bartlett2007} 
with single, double and scaled perturbative triple excitations, 
considered the `gold standard' in quantum chemistry, 
using the F12a correction~\cite{
adler2007simple,
knizia2009simplified} 
subject to an adequate scaling of the triples~\cite{knizia2009simplified}
together with the aug-cc-pVTZ basis set~\cite{
kendall1992electron,
woon1994gaussian}.
The counterpoise (cp) 
correction~\cite{
boys1970calculation} 
has been applied in order to correct for the remaining basis set superposition error, 
thus providing the \mbox{CCSD(T$^*$)-F12a/aVTZcp} technique
which we abbreviate simply by `CC' for brevity,
see also the Supplementary Information.
All interaction energies have been computed with the 
\texttt{Molpro} 
program~\cite{
molproprogram,
molpro-review}.

In order to generate the AHR-averaged NNP for the interaction, it
is firstly needed to evaluate the averaged interaction energy.
This AHR average is performed using an in-house program coupled to 
the \texttt{CP2K} program~\cite{
cp2k}, 
which is used to evaluate the NNP energies of the all-atom system. 
In a first step, the Hamiltonian matrix 
\begin{equation}
\langle \psi |\ \hat{T}_{\phm} + \frac{\hbar^2}{2\mu R^2} \hat{j}^2_{\phm} + 
V(R,\theta,\phi,\theta^\prime,\phi^\prime)\ | \psi \rangle\ 
\label{eq: hamiltonian matrix}
\end{equation} 
is computed using a linear combination of spherical harmonics
up to $l_{\rm max}=8$ to represent the eigenstates $\psi$
as specified in Eq.~(\ref{eq: lc spherical harmonics}). 
Importantly, only even values of $l$ together with the corresponding $m$~values 
are included in the sum as required to describe \ph{}. 
The potential energy 
contribution according to 
Eq.~(\ref{eq: integration potential term})
is integrated numerically using 32~Gauss-Legendre and 
64~Gauss-Chebyshev 
quadrature points for the integration of $\theta^\prime$ and 
$\phi^\prime$ angles, respectively, 
where the 
all-atom 
$V(R,\theta,\phi,\theta^\prime,\phi^\prime)$ 
potential is represented
using the all-atom NNP. 
Then, the Hamiltonian matrix is diagonalized 
in that basis 
and its lowest eigenvalue 
is the AHR-averaged energy, $V^{\text{AHR}}(R, \theta,\phi)$.
More details on the convergence of the AHR average as a function
of $l_{\rm max}$ and the number of quadrature points for the 
integration of the potential energy matrix elements can be found in the
Supplementary Information.

The AHR-averaged NNP is obtained using the same automated 
procedure as in the all-atom version
described above. 
The generation of the \ph{} center of mass positions
is done analogously as in the original procedure 
for helium 
interactions~\cite{schran2018high}
but using the same cutoffs as in the 
all-atom case
as specified above. 
But now, our in-house AHR-averaging program is used
in combination with the all-atom NNP to compute the 
reference energies 
instead of the highly accurate, but computationally expensive
\mbox{CCSD(T$^*$)-F12a/aVTZcp}
calculations used for the generation of the all-atom NNP.
Again, it is important to point out that the generation of the 
AHR-averaged interaction NNP is started from the solute's
intramolecular 
NNP
for both, \wat{} and \pwat{} as 
published earlier~\cite{schran2020automated}
in order to fully account for the flexibility of the molecules
also when interacting with \ph{}.

The AHR-averaged NNPs for \mbox{\wat{}$\cdots$\ph{}} and \mbox{\pwat{}$\cdots$\ph{}} 
are obtained doing 60 loops of the automated procedure and 
adding 100 structures to the dataset in each loop. 
Similarly to the all-atom NNPs, the structures added
in each loop have been sampled out of a set of
200~000 randomly-generated configurations, which is reduced to
100 by our active learning procedure.
The fits 
to generate the 
AHR-averaged NNPs
have been performed using NNPs composed of 2~hidden 
layers with 20~nodes each and the dataset has been divided
again
into a 
training
set composed of 90~\% of the structures and
a test set containing the remaining 10~\%.

Finally, the committee disagreement 
between the different all-atom NNP models
has
been used to remove highly distorted configurations from the 
AHR-averaged NNP
dataset.
Indeed a few 
unrealistically
distorted configurations are selected by the automated procedure
to fit the interaction NNP
which, when included, only 
deteriorate
the fit and decrease the
overall accuracy of the resulting NNP
rather than improving its quality. 
Such exotic \mbox{\wat{}$\cdots$\h{}} or \mbox{\pwat{}$\cdots$\h{}} configurations
arise when combining a high potential energy and thus low probability
structure of the solute species~-- for instance a close to linear 
configuration of the \wat\ molecule~-- 
with a rather low interaction energy of that quasi-linear \wat{} with \h{}. 
When fitting now the interaction NNP of \wat{} with \h{}, 
only the (favorably low) interaction energy is considered,  
while the (unrealistically high) potential energy of the 
bare solute molecule (which increases enormously the total potential
energy of that \mbox{\wat{}$\cdots$\h{}} complex) does not play any role. 
Describing these situations faithfully is demanding for the interaction~NNP
(if not exhaustively sampling them, which is not useful
since the underlying solute configurations have negligible probability 
in any realistic simulation) 
and should be avoided at the outset. 
In order to 
systematically 
detect these 
unrealistic \mbox{\textit{solute}$\cdots$\textit{solvent}} configurations
in an automated way,
we use the committee disagreement 
between the 10~models that constitute the all-atom NNP.
For each 
and every
selected position of the \h{} center of mass, the 
interaction
energy 
and the associated committee disagreement are evaluated for a total 
of $32\cdot64=2048$ different orientations of the \h{} molecule during the AHR
averaging procedure.
If any of those committee disagreements exceeds a certain cutoff value
then the selected configuration is discarded.
The committee disagreement cutoffs used for the AHR-averaged NNPs 
of \mbox{\wat{}$\cdots$\ph{}} and \mbox{\pwat{}$\cdots$\ph{}} 
are 0.2 and 0.03~kcal/mol,
respectively.

In order to explicitly validate the accuracy of the AHR-averaged NNPs
for the present purpose, 
we computed the interaction potential for numerous positions of \ph{} 
on 
a cubic
Cartesian
grid 
around solute species \wat{} and \pwat{} frozen at their equilibrium structures using 
(i) the CC~reference method (i.e. \mbox{CCSD(T$^*$)-F12a/aVTZcp} 
throughout including explicit AHR averaging), 
(ii) the all-atom NNP + AHR 
and (iii) the AHR-averaged NNP.
Additionally, (iv) the AHR-averaged potential from Ref.~\citenum{
zeng2011adiabatic}
has also been used for \mbox{\wat{}$\cdots$\ph{}}
since it has been obtained using a similar electronic structure methodology,
while freezing the \wat{} configuration in its equilibrium structure.
This cubic grid approach allows for the simulation of 
\ph{} microsolvation around a frozen molecule 
in order to validate the quality of the NNP representation of the
interaction potential,
which cannot be done using 
CC~reference calculations on-the-fly as the path integral simulations proceed. 
The cubic grid is generated with 43 grid points in each dimension, 
a grid spacing of 0.28~\AA , and a 
cutoff of 2.5~\AA\ around the
center of mass of the
frozen solute molecule
within which the interaction potential is not evaluated. 
This cutoff does not induce any error in our calculations 
because the energies are highly repulsive 
at short distances and, thus, 
the simulations do not sample this part of the potential energy surface.
The use of this cutoff reduces the computational effort, but
this is still not enough to 
sufficiently reduce the computational load to carry
out the required number of coupled cluster calculations on that grid. 
Of course
we also take advantage of symmetry elements of these two molecules
at their equilibrium structures, 
which all together allows for the saving of more than 75~\%
of the grid point calculations for water, reducing the number of
calculations from almost 84~000 down to about 20~000,
and 50~\% for protonated water, so 
about 40~000 calculations
are necessary.
On top of that, the number of Gauss-Legendre and
Gauss-Chebyshev
grid points for the integration of the potential energy term in the AHR
averaging
is reduced from~64 and~32 to~20 and~10, respectively,
but
only for the generation of the validation data on that grid;
we are confident that this
reduction in quadrature accuracy 
will not have
any significant impact on 
our validation of
the microsolvation
of
\wat{} and \pwat{} with \ph{}
since
we show in the  Supplementary Information that
the AHR-averaged interaction energies are
basically converged for these reduced values.
All together, these savings make the generation of the
\mbox{CCSD(T$^*$)-F12a/aVTZcp} AHR-averaged data on a grid
computationally
affordable, although it required roughly~4
and
8~million single-point CC~energy calculations 
for the frozen H$_2$O and H$_3$O$^+$ configurations, respectively. 
All Path Integral (PI) simulations have been performed with 
the \texttt{CP2K} simulation package~\cite{
cp2k}, 
where a NNP extension for the interaction potential
has been implemented by some of us.
The PI calculations on a grid 
for the purpose of explicit application of the two AHR-averaged NNPs
have been performed using 
water and protonated water molecules frozen in their 
minimum energy structure surrounded by 1 to 30 point-like \ph{} molecules. 
A 
bosonic
PIMC sampling scheme is used for \ph{}, whose 
\mbox{\ph{}$\cdots$\ph{}} 
interactions 
in the PI action
are treated using the pair density matrix approximation~\cite{
pollock1984simulation,
ceperley1995path} 
and described by 
the Silvera-Goldman pairwise interaction potential~\cite{
silvera1978isotropic}. 
The \mbox{\textit{solute}$\cdots$\ph{}} 
interactions are evaluated 
on
the grid using nearest-neighbor 
interpolation.  
The temperature is set to 1~K and 80~beads are used to discretize 
the pair density matrix.
A total number of 40 independent PIMC walkers are 
generated
for the sampling of \ph{}
together with the worm algorithm~\cite{
boninsegni2006worm,
boninsegni2006diagrammatic}
allowing for bosonic exchange.
We refer the interested reader to Ref.~\citenum{brieuc2020converged} 
where our underlying simulation technique is reviewed
including our extension of worm sampling.

\section{Results and discussion}
To illustrate the capabilities of our methodology,
we have applied it to the \mbox{\wat{}$\cdots$\ph{}} and
\mbox{\pwat{}$\cdots$\ph{}} interactions.
We chose \mbox{\wat{}$\cdots$\ph{}} since it is the same system as
studied in the original work 
introducing 
the AHR 
averaging 
procedure~\cite{
zeng2011adiabatic}.  
This makes 
the water molecule
particularly useful for us as a benchmark
to assess the accuracy of our approach. 
While the original AHR procedure relied on the rigid rotor
approximation for the impurity~--- being very reasonable
for quasi-rigid species such as the \wat{} molecule~---
our approach includes the full flexibility of the solute of choice.
For these reasons, we have also developed an AHR-averaged NNP
for \mbox{\pwat{}$\cdots$\ph{}}, as the hydronium cation is governed
by large-amplitude umbrella inversion motion. 
Furthermore, the application to 
the more weakly and strongly interacting dopants \wat{} and \pwat{}, respectively, 
while fully accounting
for stiff intramolecular flexibility as well as floppiness 
highlights the general nature of our data-driven procedure.

\subsection{\wat{}$\cdots$\ph{} interaction
and microsolvation}

We start by applying the methodology described
above to the \mbox{\wat{}$\cdots$\ph{}} interaction potential. 
The main reason for selecting this system
is to benchmark our method against the
already available AHR-averaged interaction potential~\cite{
zeng2011adiabatic},
which is based on a 
5-dimensional interaction potential 
of a rigid water molecule 
fitted 
to
accurate coupled cluster reference calculations~\cite{
valiron2008r12}.
The electronic structure setup used for these reference calculations 
is very comparable to the more recent \mbox{CCSD(T$^*$)-F12a/aVTZcp} technique 
underlying our development of the all-atom interaction NNP.
This makes the water system the perfect candidate
to serve as  
benchmark of our AHR-averaged interaction NNP.

\begin{figure*}[]
\centering
\includegraphics[width=\textwidth]{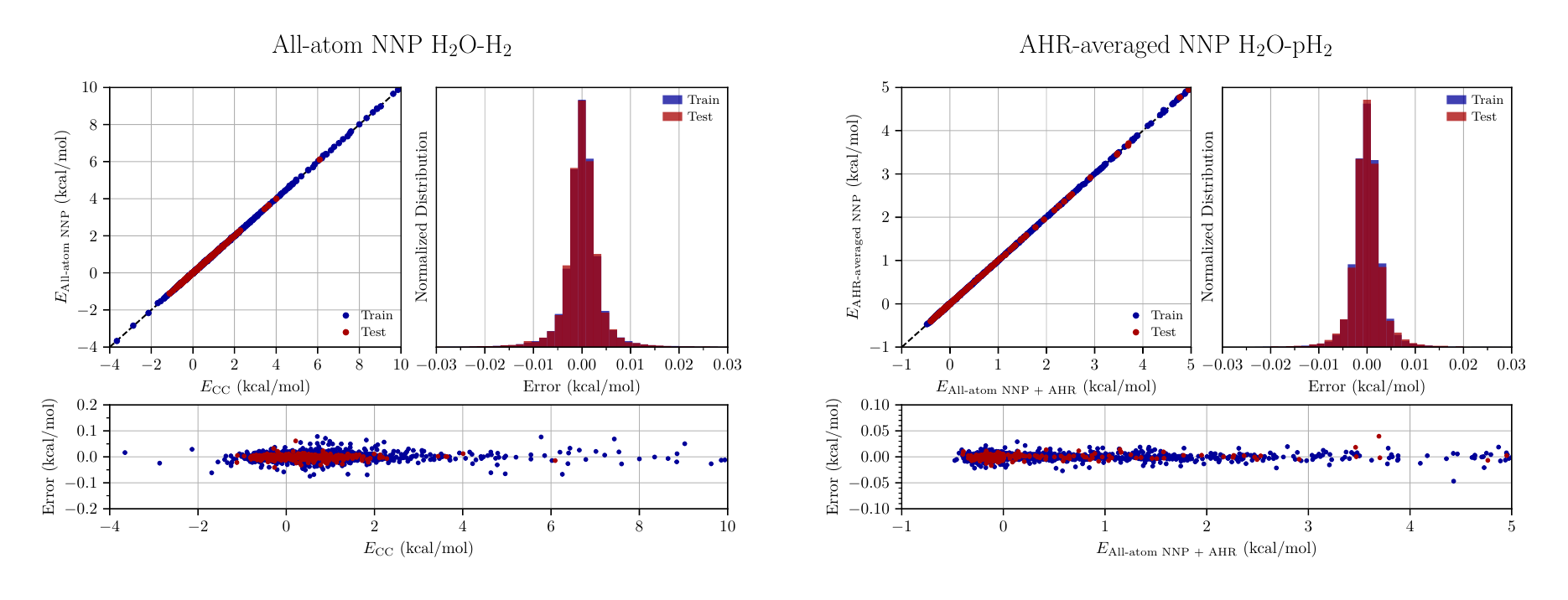}
\caption{
Accuracy of the all-atom 
\mbox{\wat{}$\cdots$\h{}} 
(left) 
and AHR-averaged \mbox{\wat{}$\cdots$\ph{}} (right) NNPs, 
see text for discussion. 
Reference energies have been computed using \mbox{CCSD(T$^*$)-F12a/aVTZcp}
for the all-atom NNP,
while the AHR-averaged energies based on the all-atom NNP 
are used as reference for the AHR-averaged NNP.
} 
\label{fig: water correlation plots}
\end{figure*}

The all-atom NNP 
underlying the generation of 
our final 
\mbox{\wat{}$\cdots$\ph{}} interaction
potential 
has been trained on a dataset containing 150~768 structures
that have been 
selected during our automated and data-driven machine learning
protocol.
This interaction potential
has a training error of $1.4\times 10^{-2}$~kcal/mol and 
a 
similar
test error of $1.4\times 10^{-2}$~kcal/mol, 
which is well below what 
is considered to correspond to chemical
accuracy (1~kcal/mol).
The accuracy of our all-atom interaction potential is highlighted in 
the left panel of Figure~\ref{fig: water correlation plots}
by different measures based on the comparison of CC~reference interaction energies
with the all-atom NNP prediction.
The nearly perfect correlation of the interaction energies showcases 
the robust description of both, the attractive and repulsive regions 
of the interaction potential.
The errors over the full range of interactions (left, bottom) reveals no outlier,
while the error distribution features a sharp peak very close to zero with
tails that barely cross 0.01~kcal/mol.
We refer the interested reader to the Supplementary Information for
additional tests of the all-atom NNP, 
including angular and radial scans 
that explicitly probe distinctly different 
regions of this interaction potential.

Using the carefully benchmarked all-atom NNP,
we proceeded to generate the AHR-averaged NNP 
to describe the \mbox{\wat{}$\cdots$\ph{}} interaction.
The AHR-averaged NNP 
relies on
a dataset containing 59~933 structures,
again selected using our automated machine learning procedure.
The fit
is subject to 
a training and a test error of $3.5\times 10^{-3}$
and $3.9\times 10^{-3}$~kcal/mol, respectively.
The accuracy of the AHR-averaged interaction energies is
illustrated in Figure~\ref{fig: water correlation plots} (right).
We find close to perfect correlation
between the AHR-energies computed with the all-atom NNP,
which provides the reference data in this case, 
and the AHR-averaged NNP prediction and very narrow
error distributions, while no outliers are observed over the
full range of interactions.
We mention in passing that the committee-based approach
to identify unrealistically distorted solute configurations 
within the 
\mbox{\textit{solute}$\cdots$\h{}} 
complexes 
as described in Section~\ref{sec:compdet} turned 
out to be mandatory in order to achieve such high quality fits.

Let us next concentrate on an in-depth benchmark of the interaction potentials.
Usually, this would entail to perform the desired simulations
to probe various properties of the system of interest
with the developed models as well as 
explicitly using 
the reference method
on-the-fly 
as customarily done when using computationally economical
DFT-based electronic structure to provide the reference data.
However, this would require the computation of the AHR-average using 
the reference \mbox{CCSD(T$^*$)-F12a/aVTZcp} method
directly.
Unfortunately, this is out of scope, as roughly 2000
such CC~calculations are needed
for each center of mass position of the \ph{} species to sample the underlying \h{} orientations 
for each and every \mbox{\textit{solute}$\cdots$\ph{}} configuration 
generated
during the path integral simulation, which easily extends to millions of such single-point calculations
for a typical HPIMD/MC simulation.
A more realistic option is to compare radial and angular scans or two-dimensional cuts 
of the full-dimensional interaction potential energy surface using the all-atom NNP
to evaluate the AHR-average
along these scans and cuts as we indeed 
show in the Supplementary Information.
However, this only allows for the testing of a small 
part
of the relevant configuration space
and, moreover, does not consider the 
indirect impact of the 
\mbox{\ph{}$\cdots$\ph{}} interactions
which greatly interplay with the \mbox{\textit{solute}$\cdots$\ph{}} interactions
.
Recall
that the spatial structure of quantum clusters at low temperature can be distinctly 
different from their structure in global and even low-energy local minima 
of their potential energy surface 
due to the severe quantum delocalization of light particles such as \ph{} or also He. 
Taking these considerations into account, 
we resort to another approach as originally introduced in Ref.~\citenum{
kuchenbecker2017constructing}
and successfully applied in the context of helium (micro)solvation~\cite{schran2018high}.
By evaluating the interaction potential for a fixed configuration of the solute
on a relatively coarse grid
using the CC~reference method, 
the \mbox{\textit{solute}$\cdots$\ph{}} interaction potential
can efficiently be mapped, 
enabling the evaluation of the 
interaction energy
on that grid during exhaustive PIMC sampling of the \ph{} solvent (while using 
nearest-neighbor interpolation
between the grid points). 
This method allows us to study microsolvation of the solute in a frozen configuration
using different numbers of \ph{} molecules.
From those calculations we can compute the spatial distribution function (SDF), 
a property that is 
known to be
highly sensitive to small changes in the interaction potential
in particular for quantum clusters for said reasons. 
Furthermore, it is possible to use this approach for both, the all-atom
and the AHR-averaged NNP,
providing a consistent benchmark over the entire process of the
generation of the AHR-averaged NNP
starting from the CC~reference calcuations. 
We 
stress 
that these 
unusually stringent
benchmarks still remain unpleasantly expensive
since they require
roughly 4 to 8~million CC~energy calculations
for the selected frozen solute configurations.
Note that this drastically exceeds
the number of CC~calculations performed
for the development of the NNPs
in the first place, 
which grants access to fully flexible solute species in HPIMD/MC simulations.

\begin{figure}[]
\centering
\includegraphics[width=0.48\textwidth]{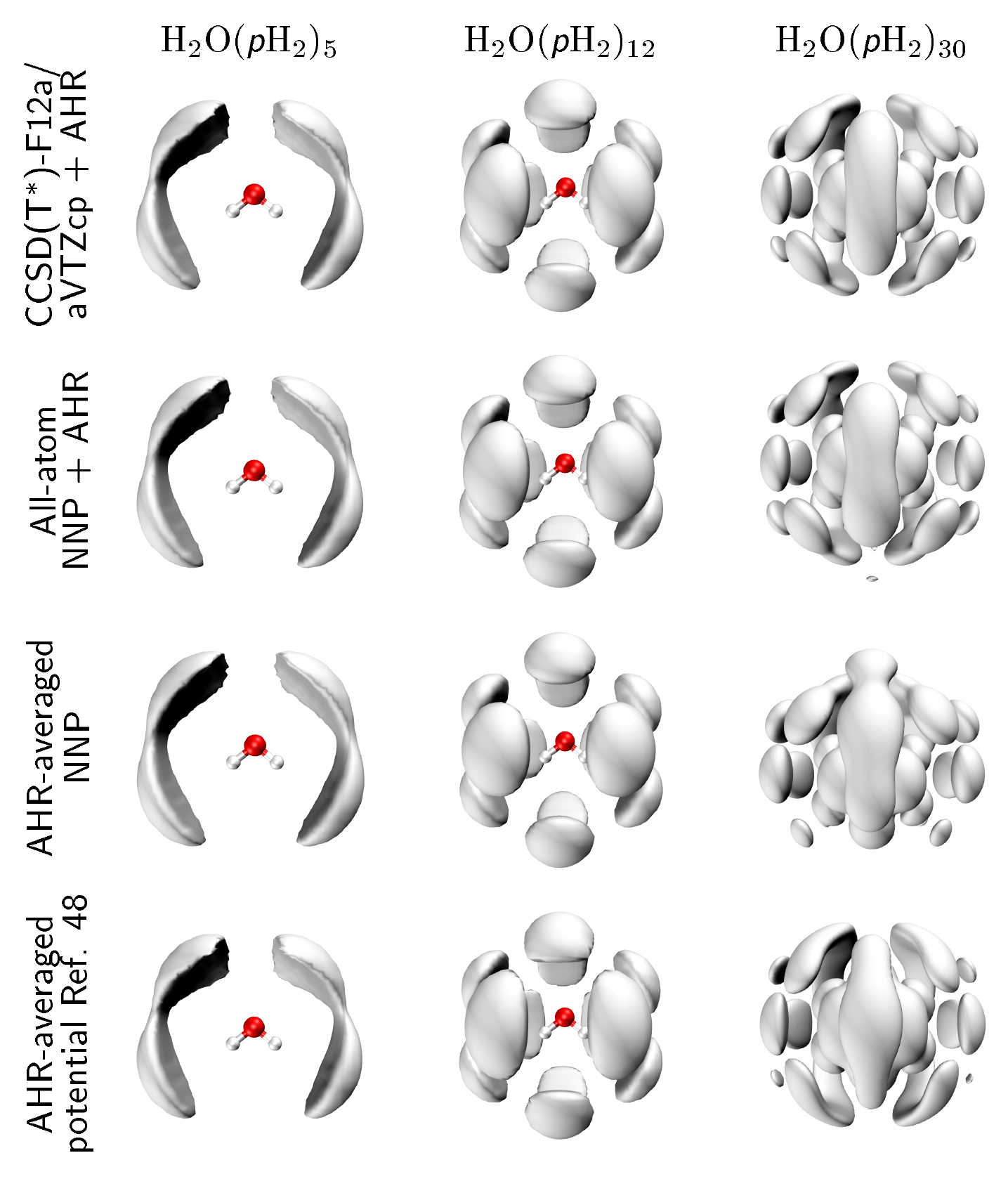}
\caption{
Comparison of the 
\ph{} SDFs around frozen \wat{} on a grid, see text, 
obtained with different methods to compute the interactions
for three 
cluster sizes $N$.
From top to bottom:
(i) \mbox{CCSD(T$^*$)-F12a/aVTZcp} + AHR-averaging,
(ii) all-atom \mbox{\wat{}$\cdots$\ph{}} NNP + AHR averaging,
(iii) AHR-averaged \mbox{\wat{}$\cdots$\ph{}} NNP,
and 
(iv) AHR-averaged potential from Ref.~\citenum{
zeng2011adiabatic}. 
From left to right: $N=5$, 12 and 30 \ph{} molecules 
solvating the minimum energy structure of the water molecule. 
The isovalues are 0.0065, 0.0090, and 0.0030~1/bohr$^3$ for $N=5$, 12 and 30.
}
\label{fig: wat ph2 grid densities}
\end{figure}

The SDFs for \mbox{\wat{}(\ph{})$_N$} have been computed from path integral
simulations using $N=1$~up to $30$~\ph{} molecules
with the interaction energy computed in four different ways
as depicted from top to bottom in Figure~\ref{fig: wat ph2 grid densities}:
(i)
using the reference CC~method and performing the AHR-average explicitly
based on single-point CC~calculations to sample
the
orientations at each center of mass position,
(ii) performing the AHR-average but using the all-atom NNP interaction energies,
(iii) using directly 
the AHR-averaged NNP,
and 
(iv) with the published AHR-averaged potential from Ref.~\citenum{
zeng2011adiabatic}.
The comparison of the obtained SDFs allows for the analysis of the
quality of the all-atom and the AHR-averaged 
NNP 
fits with respect to the highly accurate CC~reference calculations as well as to the
AHR-averaged potential available in the literature~\cite{zeng2011adiabatic}.
The results for selected clusters with $N=5$, $12$ and $30$~\ph{} are shown in
Figure~\ref{fig: wat ph2 grid densities}
from left to right,
while 
other
SDFs are presented in the Supplementary Information.
Overall, this stringent test highlights the 
high 
quality of the generated interaction
potentials as there are no significant deviations between 
the SDFs obtained with the four methods.
Only for the larger cluster size some slight 
differences can be observed
with minor variations in the population of the second solvation shell.
This has already been seen and discussed in a previous study for the 
case of 
helium solvation of protonated water clusters~\cite{schran2018high}
and traced back 
to the high sensitivity of the SDFs to the chosen isovalues
for large $N$ where the density modulations in 
three-dimensional space fade out
in conjunction with using a rather coarse Cartesian grid
with a spacing of 0.28~\AA . 
For $N=30$, the
slightly larger population of the interconnections
between most populated regions in the second solvation shell
reveals that the all-atom and
the AHR-averaged NNPs have slightly higher interaction energies
than the CC~reference in those regions, 
whereas the AHR-averaged potential from  Ref.~\citenum{
zeng2011adiabatic} 
shows a shallower energy surface
compared to our CC~benchmark. 
Despite these small variations, the overwhelming agreement 
between the different approaches validates the accurate nature of our approach
thus providing overall \mbox{CCSD(T$^*$)-F12a/aVTZcp} quality for such quantum simulations.

\subsection{\pwat{}$\cdots$\ph{} interaction
and microsolvation} 

We move on to apply the developed methodology to the more challenging
case of the \mbox{\pwat{}$\cdots$\ph{}} interaction potential.
The positive charge of this solute gives rise to significantly
stronger and more anisotropic interactions with the solvent.
Furthermore, the solute is governed by its large-amplitude
umbrella inversion
whereby the three protons move across the coplanar 
transition state on the potential energy surface. 
Therefore, in this case and 
for any 
other species subject to
large-amplitude motion, it is of crucial importance to include the flexibility
of the molecule in the interaction potential~-- as done in our data-driven approach~--
to study its solvation by \ph{} or other weakly interacting solvents.

\begin{figure*}[]
\centering
\includegraphics[width=\textwidth]{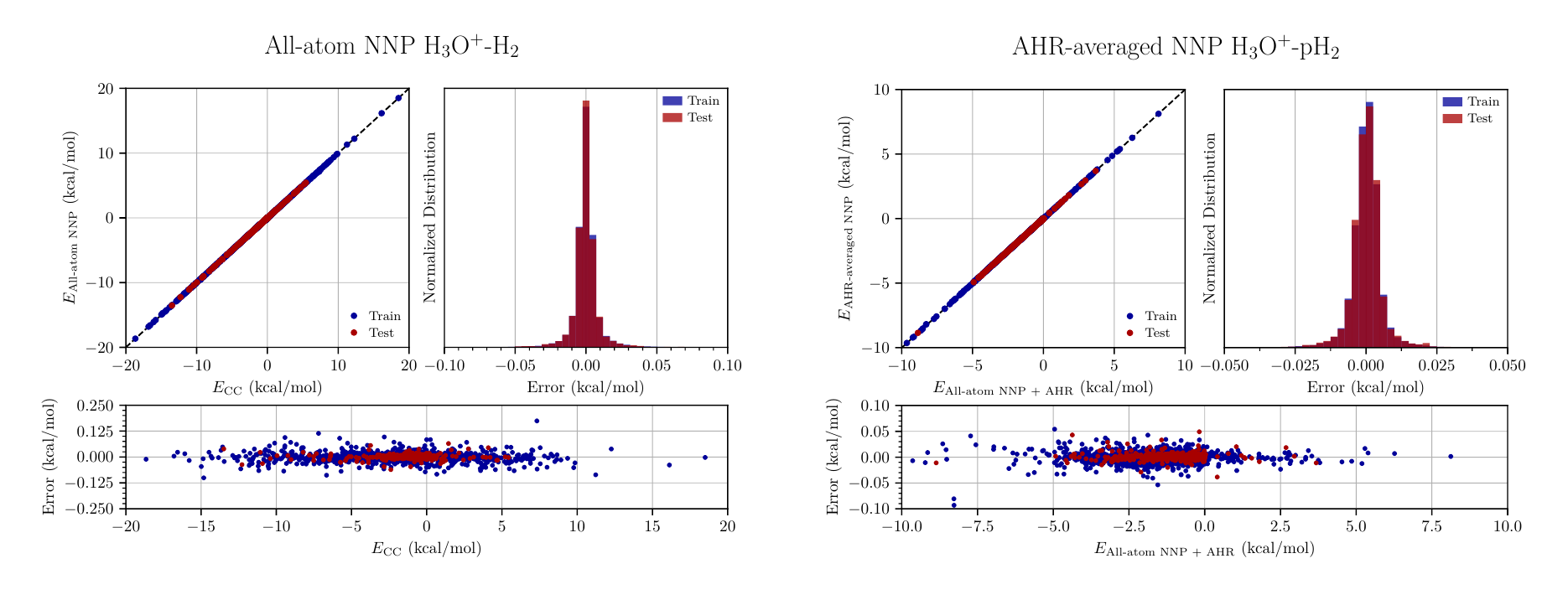}
\caption{
Accuracy of the all-atom 
\mbox{\pwat{}$\cdots$\h{}}
(left)
and AHR-averaged \mbox{\pwat{}$\cdots$\ph{}} (right) NNPs,
see text for discussion.
Reference energies have been computed using
\mbox{CCSD(T$^*$)-F12a/aVTZcp} for the all-atom NNP, while the AHR-averaged
energies based on the all-atom NNP are used as reference for the
AHR-averaged NNP.
} 
\label{fig: protonated water correlation plots}
\end{figure*}

As before, the first step in our approach is the development of the 
all-atom NNP.
This model has been trained on a dataset of 59~800 structures,
again selected in our automated procedure
.
It might appear surprising at first sight that much less
reference data were needed for \pwat{} compared to \wat{} but
the reason is that the much more shallow interaction
potential energy surface in the latter case needs to be
sampled much more densely to train the all-atom NNP
to the desired accuracy. 
The resulting model has a training and a test error of $6.5\times 10^{-2}$
and $6.2\times 10^{-2}$~kcal/mol respectively.
The quality of the training of the full dimensional interaction potential
is illustrated in Figure~\ref{fig: protonated water correlation plots}~(left),
which shows a near perfect correlation and together with a very narrow error
distribution.

In the next step, the AHR-averaged NNP has been trained
using a dataset containing
57~616 structures, 
selected as before with our automated procedure.
The resulting training and test errors are $6.0\times10^{-3}$
and $6.4\times10^{-3}$~kcal/mol, respectively.
The accuracy of this fit is 
of
the same order of magnitude 
as that of 
the AHR-averaged \mbox{\wat{}$\cdots$\ph{}} fit and it is again well below chemical
accuracy.
This very good accuracy is demonstrated in 
Figure~\ref{fig: protonated water correlation plots}~(right),
where almost perfect correlation between the AHR-averaged energies
computed a posteriori from the all-atom NNP and the predicted energies
from the AHR-averaged NNP is seen.
Furthermore, no outlier can be observed and the error distribution is
very narrow with its tail being negligible for errors larger than
$0.025$~kcal/mol.
Again, it was decisive to remove the strongly distorted 
solute configurations when fitting 
its interaction potential with \ph{}. 

We again resort to the comparison of \ph{} SDFs
for an in-depth benchmark of the performance of our models in
path integral simulations using $N=1$ up to $30$~\ph{} molecules
around frozen \pwat{}.
These SDFs are used to validate the accuracy of the NNP
at the three different stages of our approach:
(i) 
using the reference \mbox{CCSD(T$^*$)-F12a/aVTZcp} method and performing the 
AHR-average with the obtained energies,
(ii)
performing the AHR-average using the all-atom NNP,
and 
(iii)
with the AHR-averaged NNP.
As this approach requires to use the 
solute with a fixed structure,
we are considering the protonated water molecule 
in its minimum energy configuration
for that purpose. 
The results for the clusters with $N=5$, $12$ and $30$~\ph{} are shown in
Figure~\ref{fig: pwat ph2 grid densities}
whereas more are compiled in the Supplementary Information.
Overall, there are no significant deviations between the three methods,
except for some small differences similar to what has been observed for water,
again factoring in the resolution due to a grid spacing of 0.28~\AA .
This highlights 
also for this case
the high quality of the generated AHR-averaged NNP.

%
%
%
%
%
%
%
%
\begin{comment}
If we compare the positions of the populated regions
between water and protonated water, we can see very different behavior.
%
One of the most striking differences is the position of the
volumes of accumulated density for a low number of \ph{} molecules.
%
For water, the \ph{} molecules position themselves in the same plane as 
the water molecule, mostly between the hydrogen and oxygen atoms,
leaving gaps on top of the oxygen atom and between the two hydrogen atoms.
%
In contrast, for protonated water they form a torus-like structure around the cation.
%
For a higher number of \ph{} molecules, different solvation patterns
are found as well.
%
The \ph{} molecules position themselves in a spherical, very symmetrical
distribution around water, which is the opposite for the protonated water cation.
%
In the latter case, one can observe very distinctive spatial distributions
for the \ph{} molecules, highlighting the importance of
the orientational dependence of the solvation.
%
\end{comment}

\begin{figure}[]
\centering
\includegraphics[width=0.48\textwidth]{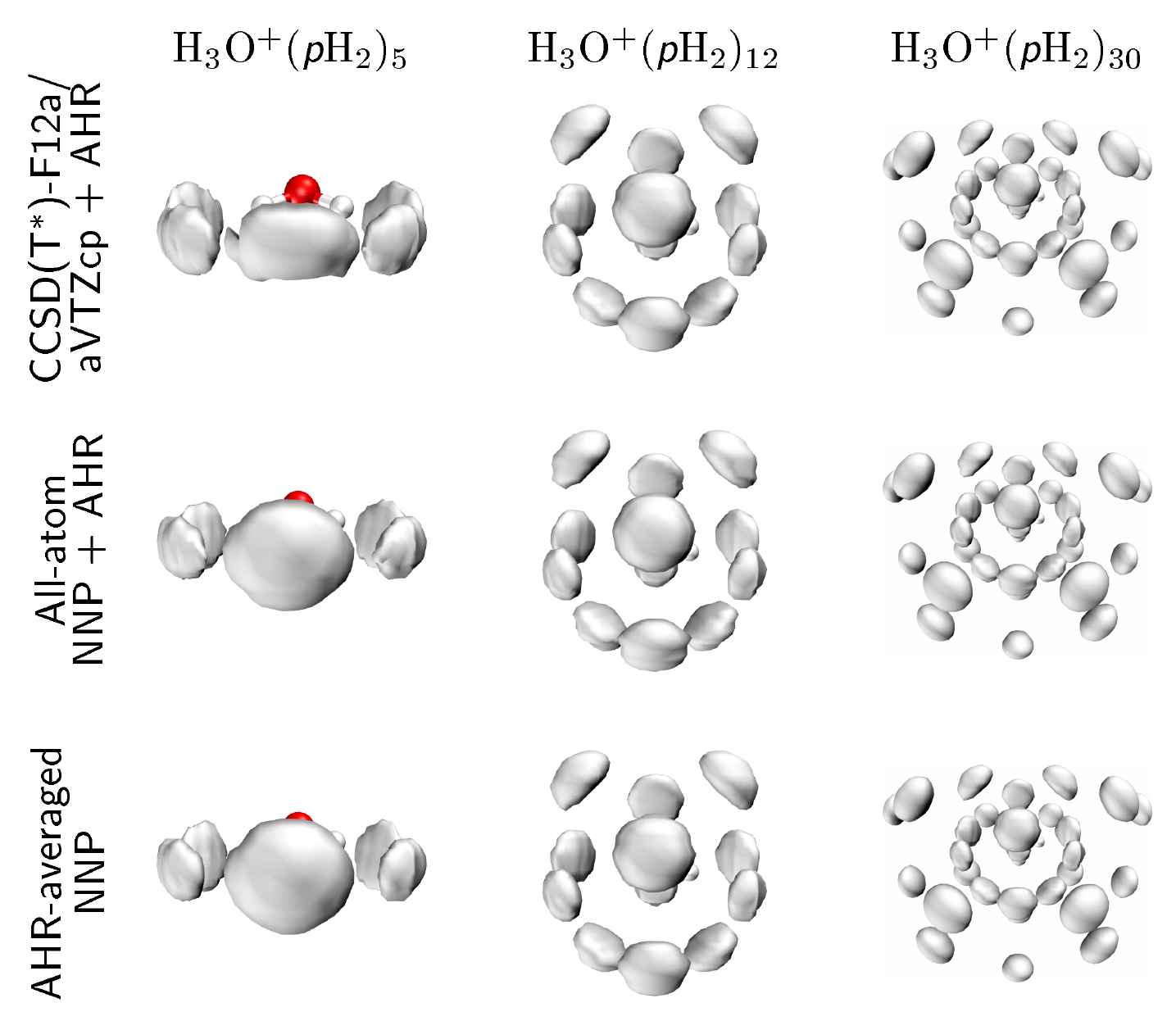}
\caption{
Comparison of the 
\ph{} SDFs around frozen \pwat\ on a grid, see text, 
obtained with different methods to compute the interactions
for three 
cluster sizes $N$.
From top to bottom:
(i) \mbox{CCSD(T$^*$)-F12a/aVTZcp} + AHR-averaging,
(ii) all-atom \mbox{\pwat{}$\cdots$\ph{}} NNP + AHR averaging,
and
(iii) AHR-averaged \mbox{\pwat{}$\cdots$\ph{}} NNP. 
From left to right: $N=5$, 12 and 30 \ph{} molecules 
solvating the minimum energy structure of the protonated water molecule. 
The isovalues are 0.0250~1/bohr$^3$ for the three 
complexes. 
}
\label{fig: pwat ph2 grid densities}
\end{figure}

\begin{figure}[]
\centering
\includegraphics[width=0.48\textwidth]{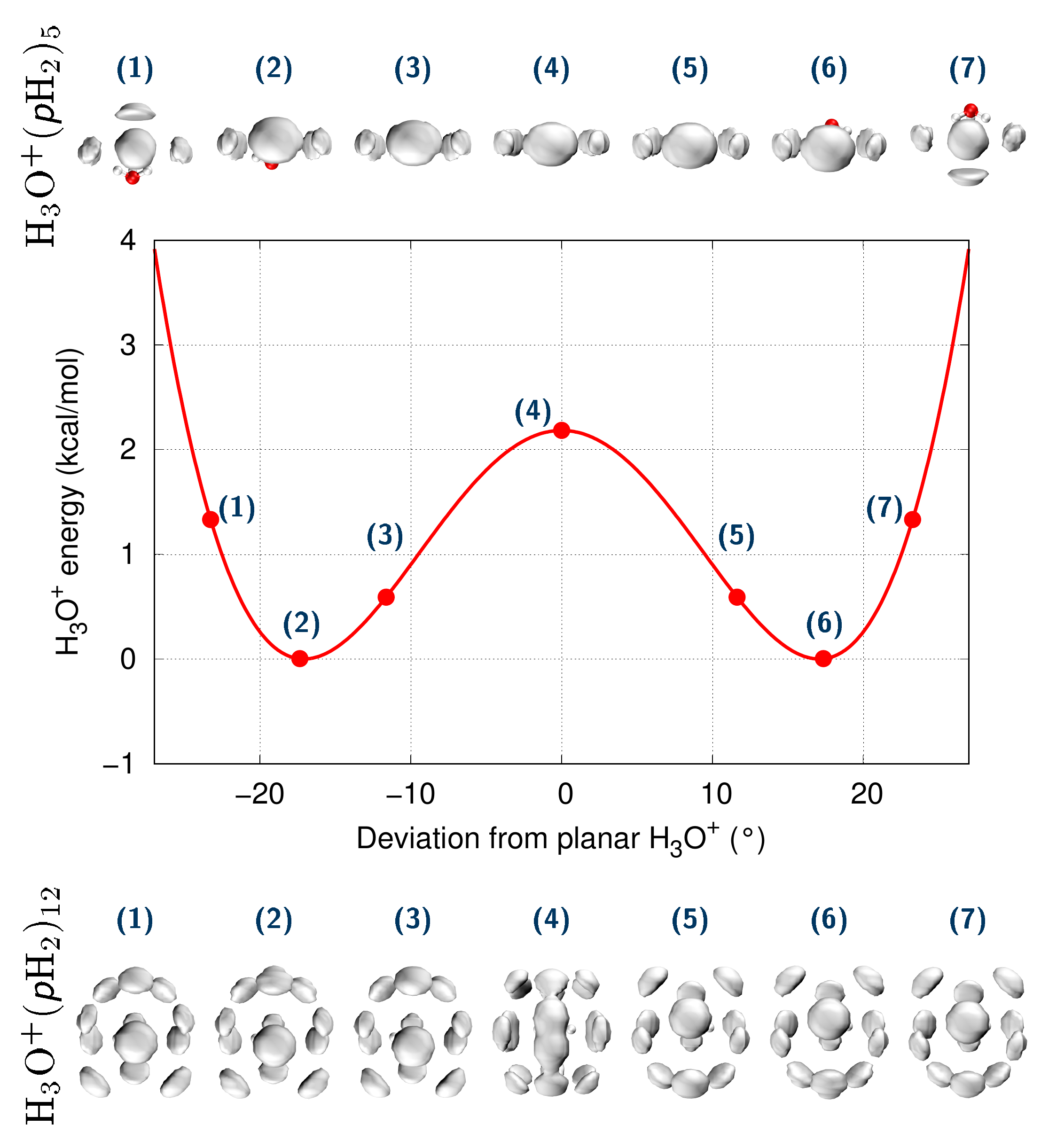}
\caption{
Comparison of the 
\ph{} SDFs around \pwat{}
on a grid obtained with the AHR-averaged 
NNP for different protonated water structures along the umbrella 
inversion
coordinate of bare \pwat{}, 
defined as the dihedral angle of the four atoms, 
which provides the deviation from planarity of the molecule; 
note that the OH bond distances are kept at
the value of the minimum energy configuration,
see text, 
for $N=5$ (top) and 12 (bottom) \ph{} molecules
together with the corresponding intramolecular potential energy profile of bare \pwat{}
in the middle panel. 
The SDF isovalues are 0.0250~1/bohr$^3$ for the two cluster sizes.
}
\label{fig: pwat scan ph2 grid densities}
\end{figure}

Finally, we illustrate the importance of including the flexibility
of the protonated water molecule in our data-driven generation of the interaction potentials by
computing the \ph{} SDFs with the AHR-averaged NNP along
the reaction coordinate of the large-amplitude umbrella motion.
We restrict this analysis to $N=5$ and 12 \ph{} molecules, but compute the
SDFs for different 
configurations
of \pwat{} along the umbrella inversion coordinate.
These SDFs 
are shown in Figure~\ref{fig: pwat scan ph2 grid densities} for
a set of 
configurations
along the umbrella inversion path
of the 
bare 
protonated water molecule 
together
with the associated 
potential
energy profile of
bare 
\pwat{} 
(obtained from its NNP published earlier~\cite{schran2020automated}). 
Here, this
coordinate is
defined as the dihedral angle of the four atoms, 
which provides the deviation from planarity of the molecule; 
note that the OH~bond distances are kept at
the value of the minimum energy configuration.
It can be easily seen that the structure of the \pwat{} molecule has a major 
impact on the spatial distribution of the solvating \ph{} molecule,
thus highlighting the importance of including the 
flexibility of the solvent in the generation of the 
AHR-averaged
NNP.
For 
the fully planar \pwat{} transition state-like structure of
\pwat{}(\ph{})$_5$, 
see point~(4) in the top panels of Figure~\ref{fig: pwat scan ph2 grid densities}, 
we observe a torus-like distribution of \ph{}
within the 
plane of the flat solute molecule.
Next, upon moving away from the transition state
toward either side along the energy path, the \ph{} density initially follows
the displacement of the protons while preserving the torus-like topology,
which also characterizes the equilibrium structure of the $N=5$ cluster
at the global minimum structure reached at points~(2) and~(6).
However, as the protons of \pwat{} come closer to each other
as a result of increasing the pyramidal character of \pwat{} further, 
the \ph{} torus 
gets
strongly compressed resulting in the displacement of one \ph{} molecule 
out of the ring as seen for the first and last SDFs with 
five~\ph{} 
at points~(1) and~(7).

A similar but even more complex scenario is observed with $N=12$ \ph{} molecules
around \pwat{}, 
see bottom panels. 
While the planar transition state 
at point~(1) again 
shows a fully symmetric SDF
including also the second solvation shell, 
when deviating from it, the SDF symmetry 
gets
broken:
Most of the \ph{} density gets more and more
accumulated on the side of \pwat{} where the protons are, thus 
leaving the oxygen side less solvated at points~(1) and~(7). 
We note that the potential energy difference between these
selected points (1) and~(7) w.r.t. the equilibrium structure~(2) and~(6)
is smaller than that of the transition state~(4) that governs
the umbrella dynamics of 
bare
\pwat{}. 
In other words: The qualitative changes of the first and second
microsolvation shell of these quantum clusters that we disclose
here are not high-energy and thus unrealistic phenomena at 
temperatures on the order of 1~K, 
but are expected to be highly relevant when
immersing a
fluxional
dopant such as \pwat{} in 
ultra-cold 
\ph{} environments. 
More generally speaking, this 
comparison highlights the potential of our overall approach
to elucidate the coupling between bosonic \ph{} environments
and the (ro-)vibrational motion of 
non-rigid 
molecular impurities
being subject to large-amplitude motion and floppiness, 
while it consistently also covers the limit of quasi-rigid
molecules as demonstrated here for \wat{}.

\section{Summary and conclusions}
In this work we have shown how highly accurate \mbox{\textit{impurity}$\cdots$\ph{}}
interaction potentials, 
which include the full flexibility of non-rigid molecular impurities
and implement the nuclear spin statistics of this solvent species, 
can be developed in a 
data-driven manner based on
largely automated
machine learning techniques. 
These high-dimensional neural network potentials (NNPs) 
enable converged quantum simulations of the corresponding para-hydrogen clusters, 
\mbox{\textit{impurity}(\ph)$_N$}, from the tagging limit for small~$N$
via the intermediate microsolvation regime
to finally bulk-like nanodroplet solvation for very large solvent numbers. 
We achieve this 
task in a three step process, where we
start by generating an all-atom NNP for the interaction between the
molecular solute impurity
and an \h{} molecule fitted to 
accurate coupled cluster electronic structure calculations,
here \mbox{CCSD(T$^*$)-F12a/aVTZcp.}  
Next, using this all-atom NNP, we obtain 
an effective 
point-particle description of \ph{} by
averaging over 
all orientiations
of the 
two-site
\h{} molecule.
Including only the even rotational quantum states in that anisotropic quantum average properly
accounts for 
the nuclear spin statistics of \ph{} within the adiabatic hindered rotor (AHR) approximation.
Finally,
we train the \mbox{\textit{impurity}$\cdots$\ph{}} AHR-averaged interaction potential
to these 
efficiently generated
AHR-averaged energies, resulting in 
the corresponding 
single-site AHR-averaged NNP. 
Most parts of this process are fully automated, 
thus enabling
\mbox{\`a la carte} 
generation of AHR-averaged NNPs
from quasi-rigid to highly fluxional molecular species or complexes.

With the aim of showing the capabilities of our data-driven methodology,
we have applied it to the \mbox{\wat{}$\cdots$\ph{}} and \mbox{\pwat{}$\cdots$\ph{}} interactions
to represent the class of quasi-rigid and non-rigid molecular dopants, respectively,
for which we 
analyze 
the accuracy of each step of the process.
Detailed comparison of the prediction of the developed NNP models highlights the
capability of our approach to 
retain 
coupled cluster accuracy
up to the level of the AHR-averaged NNPs. 
Going beyond energetic 
error
assessments, we also show that
the \ph{} probability densities
within the \mbox{\wat{}(\ph)$_N$} and \mbox{\pwat{}(\ph)$_N$} 
complexes 
with frozen solute impurities, 
as sampled from 
path integral simulations at 1~K including the bosonic exchange of \ph{}, 
are in very good agreement with 
the corresponding spatial distribution functions (SDFs) for different cluster sizes~$N$ 
obtained 
when explicitly using the 
coupled cluster reference method including explicit AHR-averaging. 
This makes us confident that our approach holds
great promise 
for the development of highly accurate \ph{} interaction potentials
for a wide variety of systems.

Going beyond the rigid-rotor approximation, usually employed in
the field, our method accounts for the full flexibility of the
solute of interest
from small- up to large-amplitude motion.
Although certainly not crucial for quasi-rigid impurities such as water, 
considering the \mbox{(ro-)vibrational} dynamics of the solute species
is of great importance for fluxional systems, 
which are completely dominated by large-amplitude motion, such as the studied hydronium cation.
Therefore, our methodology enables accurate investigations of \mbox{(micro-)solvation}
phenomena of complex molecular species and complexes in \ph{} 
environments from the tagging limit up to nanoscale solvation 
at ultra-low temperatures,
while fully accounting for the flexibility of the impurity
and the bosonic nature of \ph{}.

\section*{Supplementary material}

The all-atom 
(AA)
and AHR-averaged final interaction NNPs
for \wat{} and \pwat{} interacting with \h{} and \ph{}, respectively, 
are prodivded as separate 
supplementary material
data files as follows:
\\ 
\mbox{NN-IP-H2O-H2-AA-2022-V1}, \\
\mbox{NN-IP-H2O-PH2-AHR-2022-V1}, \\
\mbox{NN-IP-H3OP-H2-AA-2022-V1}, \\
\mbox{NN-IP-H3OP-PH2-AHR-2022-V1} \\
whereas additional information and analyses are provided in
a separate PDF~document.

\begin{acknowledgments}
We would like to thank Harald Forbert for many insightful discussions
on this topic
as well as for most valuable technical help. 
We are particularly grateful to 
Pierre-Nicholas Roy for providing us with the
AHR-averaged interaction potential for \mbox{\wat{}$\cdots$\ph{}} 
as published in 
Ref.~\citenum{
zeng2011adiabatic}
and 
for granting us
permission to use and publish the data
we obtained using this potential in Figure~3. 
Partially funded
by the \textit{Deutsche Forschungsgemeinschaft} 
(DFG, German Research Foundation) 
via the DFG grant MA~1547/19 to D.M.,
by the DFG under Germany's
Excellence Strategy~-- EXC~2033~-- 390677874~-- RESOLV,
and
supported by the ``Center for Solvation Science ZEMOS''
funded by the German Federal Ministry of Education and Research
and by the Ministry of Culture and Research of North Rhine-Westphalia.
C.S. acknowledges partial financial support from the
\textit{Alexander von Humboldt-Stiftung}.
The computational resources were provided by 
HPC@ZEMOS, HPC-RESOLV, and BoViLab@RUB.
\end{acknowledgments}

\section*{Author Declarations}

\subsection*{Conflict of Interest}

The authors have no conflicts to disclose

\section*{Data Availability Statement}

Data available in article or supplementary material

\end{document}

\title{SUPPORTING INFORMATION\\
Neural Network Interaction Potentials 
for \textit{para}-Hydrogen \\
with Flexible Molecules}
\author{Laura Dur\'an Caballero,$^{1}$
Christoph Schran,\textit{$^{1,\rm a)}$}
Fabien Brieuc\textit{$^{1,\rm b)}$}
and Dominik Marx$^{\ast}$\textit{$^{1}$}}
\affiliation{$^{1}$~Lehrstuhl f\"ur Theoretische Chemie, Ruhr-Universit\"at Bochum, 44780 Bochum, Germany
\\$^{\rm a)}$~Present address: Yusuf Hamied Department of Chemistry, University of Cambridge, Cambridge, CB2 1EW, UK 
\\ $^{\rm b)}$~Present address: Laboratoire Mati\`ere 
en Conditions Extr\^emes,
Universit\'e Paris-Saclay, CEA, DAM, DIF, 91297 Arpajon, France}

\date{\today}

\maketitle

\tableofcontents

\newpage

\section{Benchmarking the Electronic Structure Method}
The choice of the electronic structure method is 
an important ingredient in the quality of any generation of 
parametrized
potential energy functions, including
the presented data-driven methodology.
Here, it will define the ultimate accuracy 
limit
that can be achieved with  
the interaction neural 
network potentials (NNPs), 
both the all-atom and the AHR-averaged ones,
independently from the quality of the fits as such. 
In our previous work~\cite{
schran2018high}, 
where the 
interaction potential between a solute molecule
and helium has been obtained in the framework of NNPs
as reviewed recently~\cite{brieuc2020converged},
it has been demonstrated that essentially converged coupled cluster
electronic structure calculations~\cite{
cizek1966correlation, 
bartlett1989coupled,
bartlett2007}
can be used~--  in practice~--
in order to 
parametrize
interaction NNPs at this quality level
in a fully data-driven approach. 
Therein, the interaction energies have been computed using 
single, double and scaled perturbative triple excitations \mbox{CCSD(T$^*$)}, 
broadly considered the ‘gold standard’ in quantum chemistry, 
in conjunction with using the F12a~explicit correlation factor in the wavefunction~\cite{
adler2007simple,
knizia2009simplified} 
together with the aug-cc-pVTZ basis set~\cite{
kendall1992electron,
woon1994gaussian}, 
as well as the counterpoise (cp) correction~\cite{
boys1970calculation} 
considering the two interacting species
has been applied in order to correct for 
the basis set superposition error,
dubbed \mbox{CCSD(T$^*$)-F12a/aVTZcp} method also abbreviated by `CC'
in the present context. 
We have explicitly shown previously~\cite{schran2018high}, 
for the case of He~interactions with
protonated water clusters from the hydronium cation to larger ones, 
that the combination of the F12a~technique 
together with 
the cp~correction yields essentially converged \mbox{CCSD(T$^*$)} interaction energies
using that 
aug-cc-pVTZ 
triple-zeta basis set to expand the orbitals.
This is documented in Section~I.A of the Supplemental Material
of Ref. \citenum{schran2018high}
by explicit statistical comparison of \mbox{CCSD(T$^*$)-F12a/aVTZcp} 
and \mbox{CCSD(T$^*$)-F12a/aVQZcp} data
to the results obtained in the complete basis set (CBS) limit,
\mbox{CCSD(T$^*$)/CBS}.
Although we therefore know that the \mbox{CCSD(T$^*$)-F12a/aVTZcp} method 
shows excellent performance for computing
the interaction energies between solute molecules and helium,
additional 
benchmark calculations have been performed
in order to ensure that this method provides 
practically converged
results also in the case of interactions with \h{}.

\begin{figure}[]
\centering
\includegraphics[width=\textwidth]{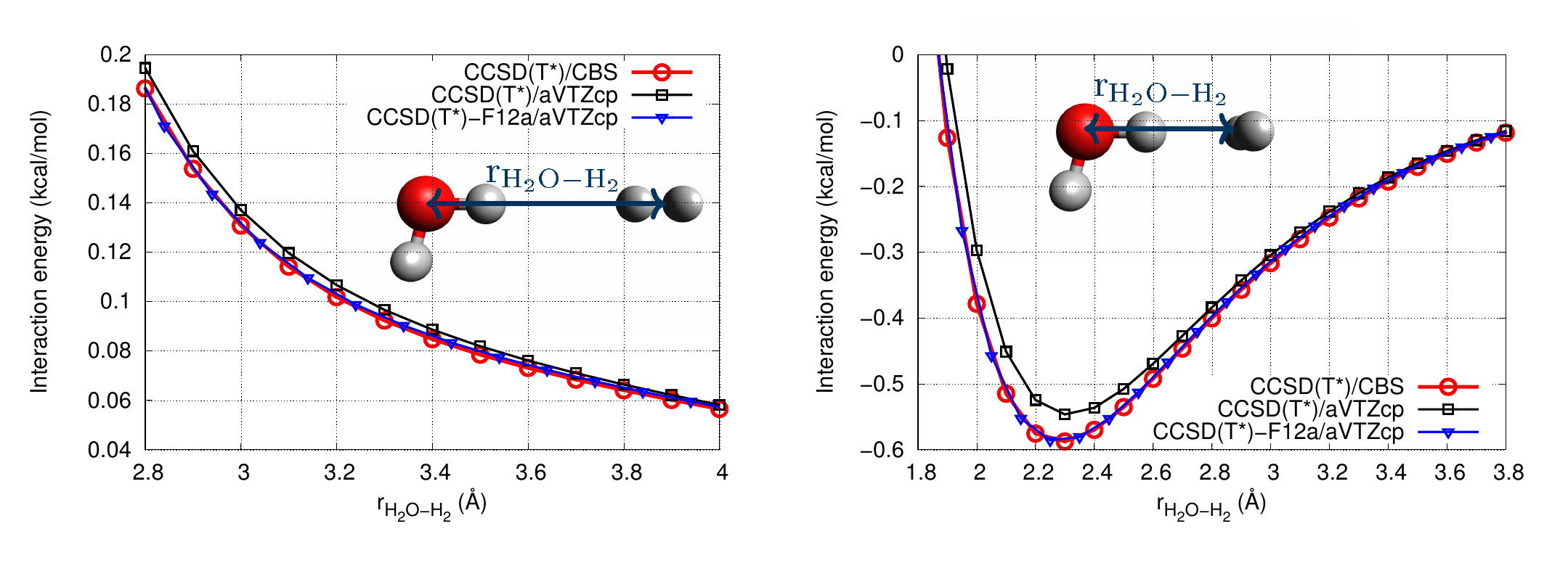}
\caption{
Results of the benchmark calculations for the \mbox{\wat{}$\cdots$\h{}}
interaction for different CC methods and basis sets, see text.
The orientation of the molecules is given as an inset,
one of them having the \h{} molecule aligned with the \mbox{O-H bond} (left),
whereas the other one has the \h{} 
oriented at $90^\circ$
with respect to the same \mbox{O-H bond} (right).
}
\label{fig: benchmark water-ph2}
\end{figure}

The electronic structure method has been benchmarked, both for the
\mbox{\wat{}$\cdots$\h{}} and \mbox{\pwat{}$\cdots$\h{}} interactions,
using 
many 
different orientations of the \h{} molecule with respect to the solute
while scanning distances.
Thus, we validate here the underlying electronic structure method by directly comparing
full potential energy profiles along relevant coordinates,
thereby 
transcending our previous purely statistical assessment~\cite{
schran2018high}.
For simplicity, only two of them for the \mbox{\wat{}$\cdots$\h{}} interaction
are shown here 
since they are representative, 
see Figure~\ref{fig: benchmark water-ph2}. 
The electronic structure methods used are the 
\mbox{CCSD(T*)/aVTZcp}, 
\mbox{CCSD(T*)-F12a/aVTZcp}, 
and \mbox{CCSD(T*)/aVQZcp} methods, 
from which the CBS limit \mbox{CCSD(T*)/CBS} has been calculated
following our earlier work~\cite{
schran2018high}.
As before,
all interaction energies have been computed in the present investigation using the
\texttt{Molpro}
program~\cite{
molproprogram,
molpro-review}.

The results for two 
important orientations 
of the \mbox{\wat{}$\cdots$\h{}} complex
are presented in \mbox{Figure~\ref{fig: benchmark water-ph2}.}
In both cases, the F12a 
explicit correlation factor 
in conjunction with using the aVTZ basis and the cp~correction provides us 
with reference interaction energies for the NNP generation 
that are indistinguishable~-- on the relevant energy scale~--
from the CBS~data.
This analysis also makes clear that this perfect agreement with the
numerically exact \mbox{CCSD(T*)} energies can only be achieved thanks to
using the F12a~factor in the expansion of the many-body
wavefunction in terms of the orbitals 
when using that triple-zeta basis set
while the traditional \mbox{CCSD(T*)/aVTZcp} method
(also including the cp~correction) consistently over/underestimates
the repulsive/attractive interactions as seen in the left/right panels. 
Very similar 
results, therefore not presented, have also been found for 
the interaction energies in 
other arrangements 
of \mbox{\wat{}$\cdots$\h{}} as well as of \mbox{\pwat{}$\cdots$\h{}}. 
The conclusion of our benchmarking is that the \mbox{(non-CBS-extrapolated)}
\mbox{CCSD(T*)-F12a/aVTZcp} interaction energies provide converged CC~reference data 
to which the interaction NNPs will be fitted, thus avoiding explicit extrapolation
based on multiple basis set calculations for each configuration in the reference data set. 
Therefore, the \mbox{CCSD(T$^*$)-F12a} method together with the aug-cc-pVTZ basis set 
and the counterpoise correction
is used to describe the  \mbox{\textit{impurity}$\cdots$\h{}} interactions.

Last but not least, it is 
important to point out that the interaction energies for one 
of the orientations shown in Figure~\ref{fig: benchmark water-ph2}
are repulsive for the whole range of distances (left), 
whereas the other orientation (right) shows an 
attractive
minimum.
This gives an indication on how anisotropic the potential is, 
as a change in the orientation of the \h{} molecule by $90^\circ$ results
in a change from the repulsive regime to a minimum in the potential energy surface
of about 0.6~kcal/mol at a distance of 2.8~{\AA}
and increases even more at smaller distances (not shown). 
An example of this would be at a distance of 2.3~{\AA}, where the interaction energy
for the scans in the left and right panels of Figure~\ref{fig: benchmark water-ph2}
are about~0.7 and $-$0.6~kcal/mol respectively, adding up to a total difference of 1.3~kcal/mol
for a rotation of the \h{} molecule that amounts to $90^\circ$.
Such energy changes are not only very relevant on the scale of the maximum interaction energy 
between two \ph\ molecules of $\approx 0.07$~kcal/mol
(according to the Silvera-Goldman effective pair potential~\cite{silvera1978isotropic}),
but also significantly affect the adiabatic hindered rotor (AHR) averaging in large regions 
of the intermolecular configuration space. 
Therefore, the use of the AHR-averaging approximation to go beyond
the standard spherical average of the interaction potential is of great importance.
In the AHR-averaging context, it is also reassuring 
to infer from Figure~\ref{fig: benchmark water-ph2}
that the \mbox{CCSD(T*)-F12a/aVTZcp} used here throughout performs consistently well
compared to the CBS~benchmark to quantify both, 
repulsive and attractive interaction energies at the same intermolecular distance.

\clearpage

\section{All-atom \wat{}$\cdots$\h{} Interaction NNP}

\begin{figure}[h!]
\centering
\includegraphics[width=\textwidth]{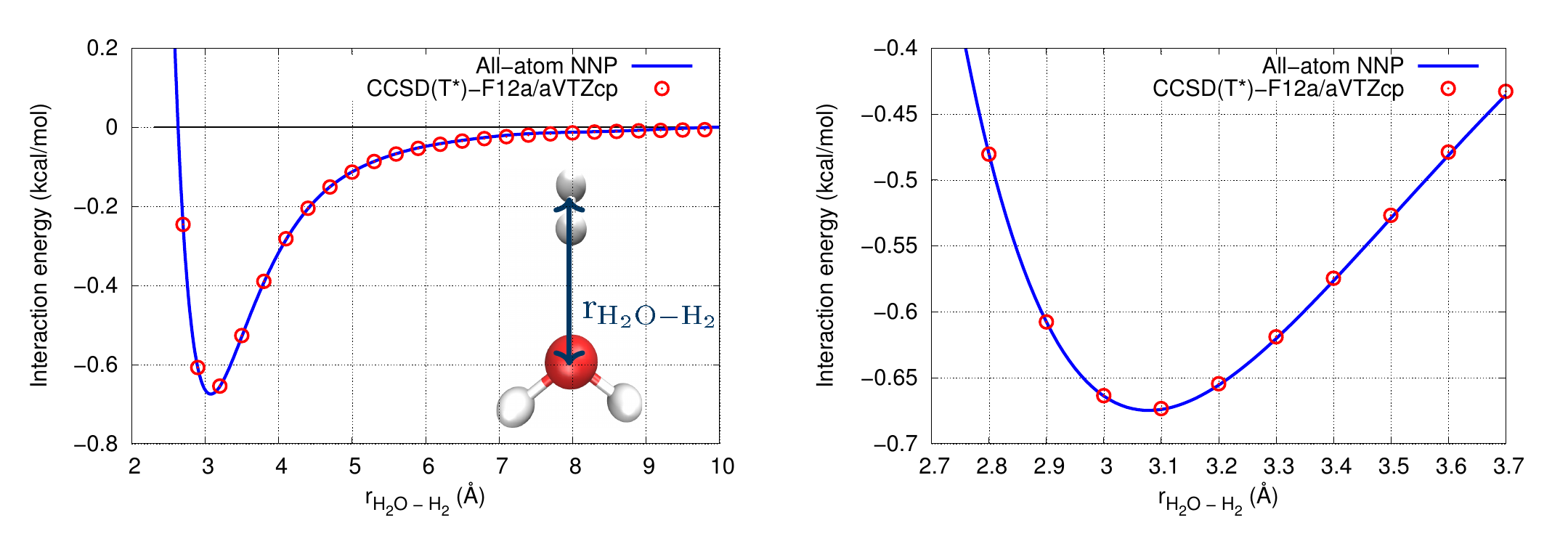}
\caption{
Comparison between the interaction energies computed with the
reference \mbox{CCSD(T$^*$)-F12a/aVTZcp} method
(symbols mark these single-point numerical data here and in subsequent such plots)
and the all-atom NNP
(solid line marks these continuous analytical data here and in subsequent such plots)
for the radial scan along the $C_{2v}$ axis
of the water molecule
with \h{} in the same plane as the \wat{} molecule (see inset).
The right panel
magnifies the region around the
minimum of the interaction energy profile. }
\label{fig: fd nnp radial h2o}
\end{figure}

\begin{figure}[]
\centering
\includegraphics[width=\textwidth]{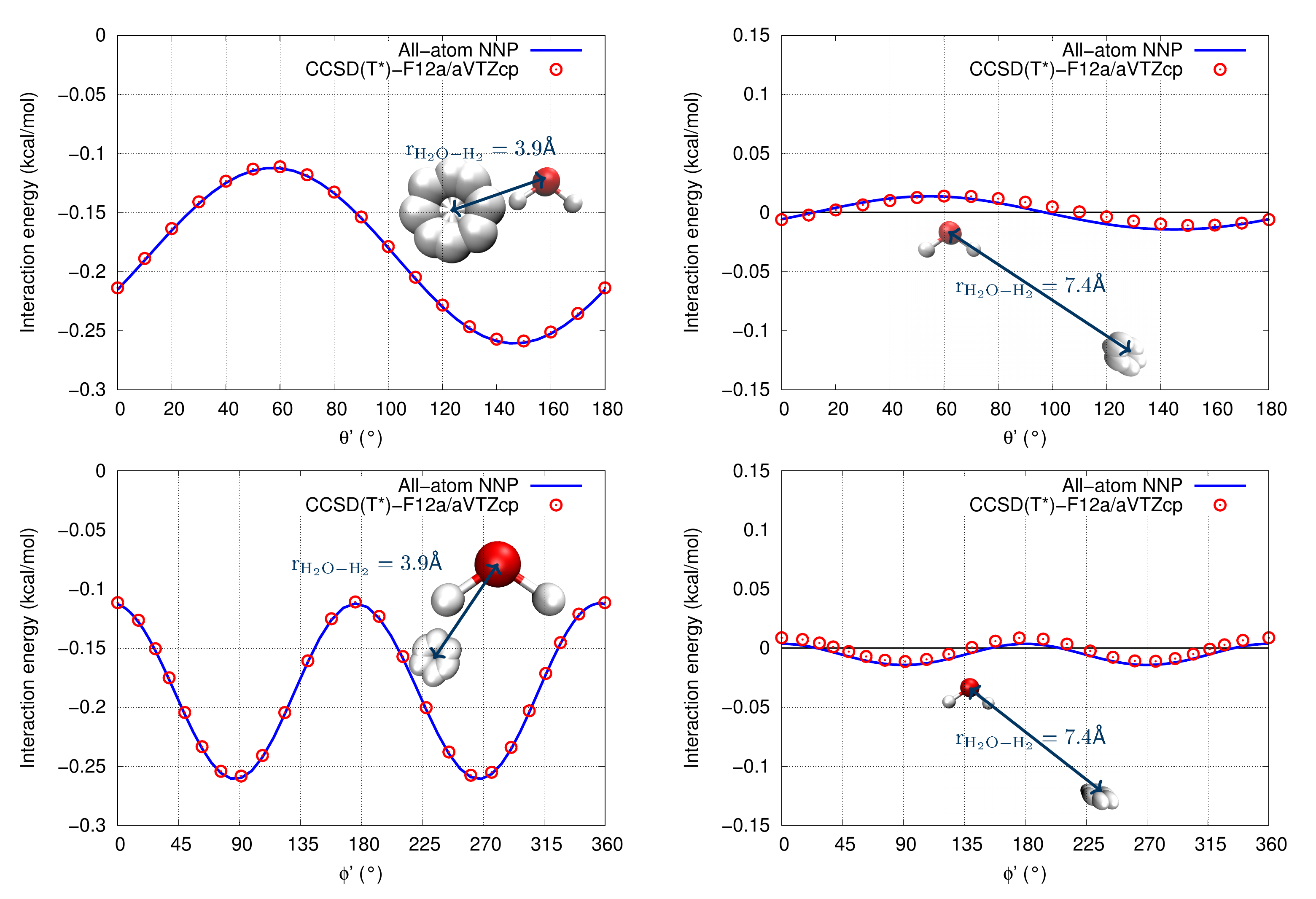}
\caption{
Comparison between the interaction energies computed with the
reference \mbox{CCSD(T$^*$)-F12a/aVTZcp} method and the all-atom NNP
for two different conformations of the \mbox{\wat{}$\cdots$\h{}} system (left and right)
as a function of the rotation of the \h{} molecule around its center of mass
along the \mbox{$\theta^\prime$ (top)} and \mbox{$\phi^\prime$ (bottom)} angles. }
\label{fig: fd nnp angular h2o}
\end{figure}

\begin{figure}[]
\centering
\includegraphics[width=\textwidth]{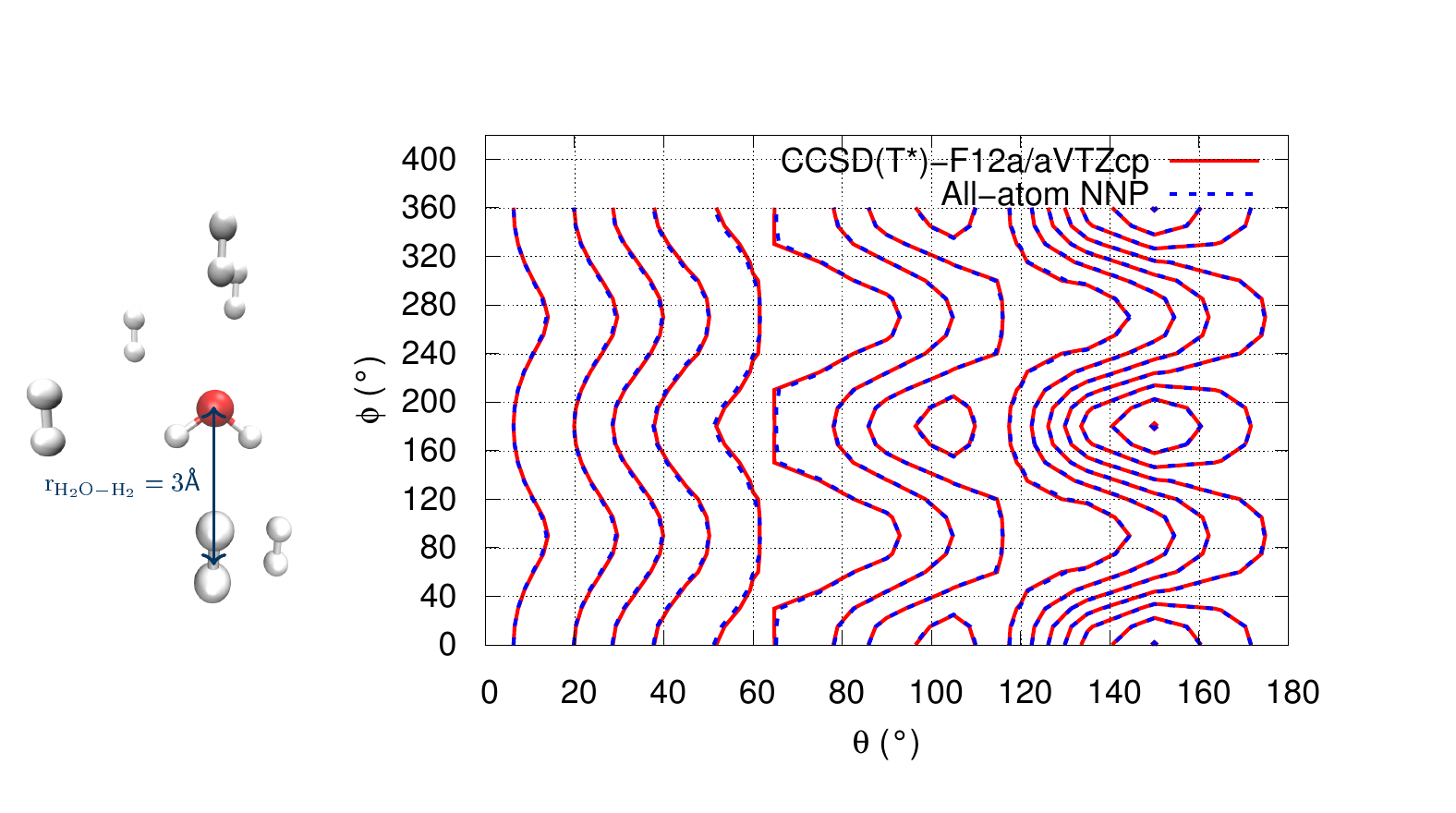}
\caption{
Comparison of the contour plots obtained with the all-atom NNP interaction energies 
and the \mbox{CCSD(T$^*$)-F12a/aVTZcp} interaction energies 
for a two-dimensional scan over the orientation of the \h{} molecule 
around the \wat{} molecule at the global minimum energy structure,
where the orientation of \h{} with respect to its center of mass
is kept fixed. 
The energy scale corresponding to this contour plot is 
$-$3.0 to 5.0~kcal/mol.
}
\label{fig: fd nnp 2d scan 1 h2o}
\end{figure}

\begin{figure}[]
\centering
\includegraphics[width=\textwidth]{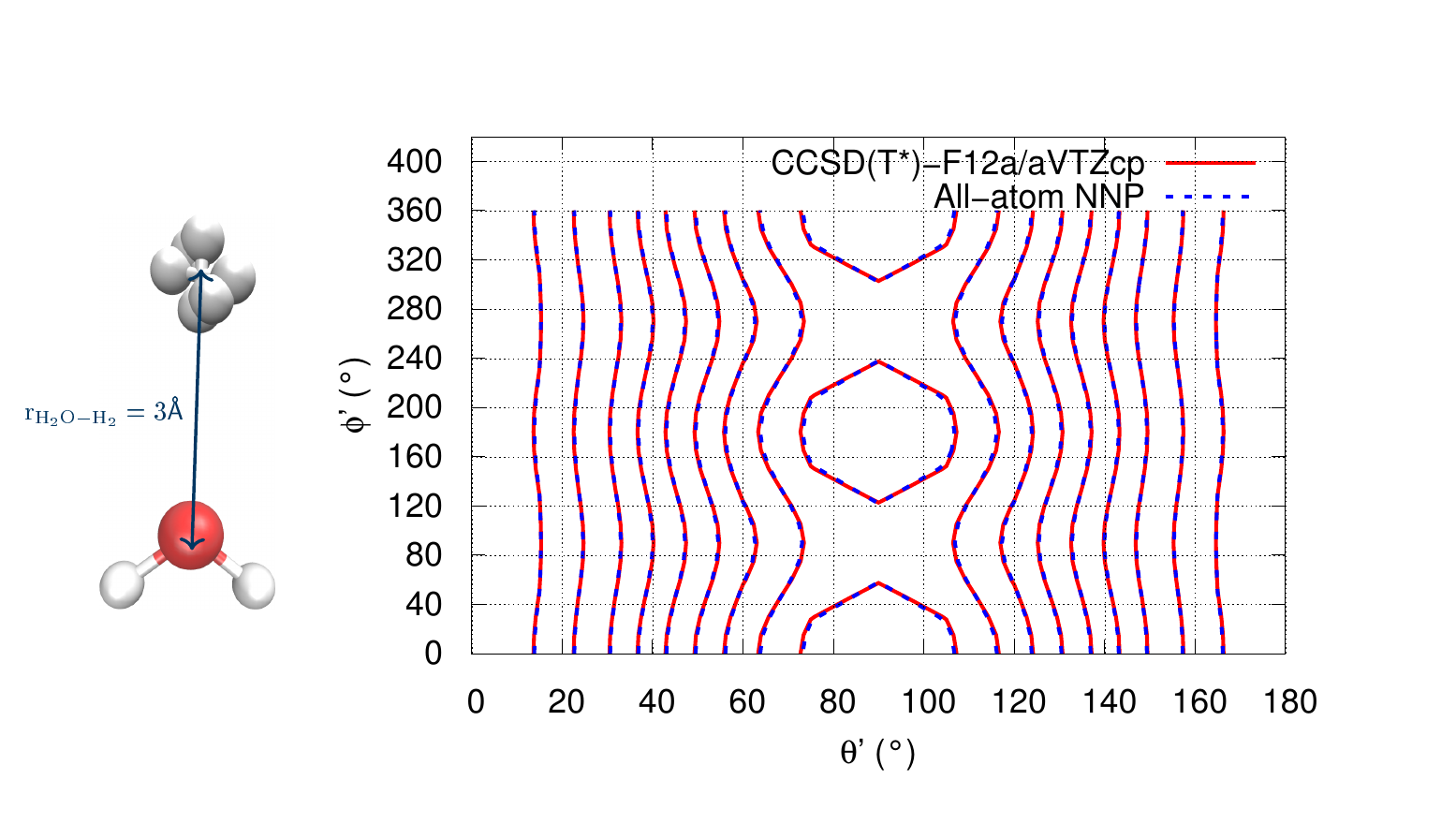}
\caption{
Comparison of the contour plots obtained with the all-atom NNP interaction energies 
and the \mbox{CCSD(T$^*$)-F12a/aVTZcp} interaction energies 
for a two-dimensional scan over the orientation of the \h{} molecule around its center of mass,
with the \wat{} molecule at the global minimum energy structure
keeping the \h{} center of mass fixed. 
The energy scale corresponding to this contour plot is 
$-$3.0 to 1.5~kcal/mol.
}
\label{fig: fd nnp 2d scan 2 h2o}
\end{figure}

The two all-atom NNPs that have been generated for the two solute molecules 
as explained in the main text need to be carefully tested 
with special emphasis on accurately describing~--
at the level of that interim~NNP~--
the rotation of the two-site \h{} molecule 
around its center of mass. 
This is especially important because the AHR-averaging procedure requires 
interaction energies for very many different orientations of the \h{} molecule
and the accuracy on those interaction energies~--
as described by the all-atom NNP~--
will translate to that of the final AHR-averaged 
NNP. 
The easiest testing of the accuracy of the 
all-atom 
fits is a direct comparison 
between the reference electronic structure method, 
\mbox{CCSD(T$^*$)-F12a/aVTZcp} here,
and the predicted interaction enegies from the generated all-atom NNP.
This has been performed for several scans along the interaction potential energy
surface, namely radial, angular and also two-dimensional cuts, 
see Figures~\ref{fig: fd nnp radial h2o}
to \ref{fig: fd nnp 2d scan 2 h2o}.

First, a radial scan for a frozen structure of the water molecule and 
fixed orientation 
of the \h{} molecule is shown in Figure~\ref{fig: fd nnp radial h2o}.
It can be observed that the quality of the all-atom NNP
along this scan is almost perfect
when compared to the explicit single-point reference electronic structure calculations.
Similar accuracy is found in Figure~\ref{fig: fd nnp angular h2o},
where the reference CC~interaction energies
and the energies computed from the all-atom NNP 
are compared for four different angular scans of \mbox{\wat{}$\cdots$\h{}}.
These 
scans correspond to two different 
arrangements of the system
in the left and right panels as depicted in the insets,
while the rotation is performed
w.r.t. 
different angles,
namely along the $\theta^\prime$ and $\phi^\prime$ angles 
in the top and bottom panels, respectively.
These one-dimensional interaction energy profiles already give a hint of 
the favorable accuracy of the all-atom 
interaction NNP for the rotation of the \h{} molecule 
around its center of mass,
but a more 
stringent benchmark is provided by 
two-dimensional cuts through the corresponding potential energy surface as follows. 

Thus, 
two-dimensional cuts of the interaction potential energy surface have been scrutinized.
One of them has a fixed distance between the \wat{} and the \h{} molecules
as well as a fixed orientation of the \h{} molecule with respect to its center of mass
whereas the relative orientation between the two molecules is changed
(see Figure~\ref{fig: fd nnp 2d scan 1 h2o}).
The second one has a fixed position of the \h{} center of mass
with respect to the water molecule whereas the orientation of the \h{} molecule
is varied 
(see Figure~\ref{fig: fd nnp 2d scan 2 h2o}).
In both cases, the comparison between the interaction energies is
shown by superimposing contour plots.
The agreement found between the CC~reference and
the all-atom NNP interaction energies is almost perfect.
The favorable comparisons of
these
two-dimensional scans are very significant,
since they show that the loss of accuracy when
performing the AHR average using the all-atom interaction NNP 
will be
minimal when compared to the 
computationally very demanding
\mbox{CCSD(T$^*$)-F12a/aVTZcp} method.
This almost perfect accuracy is mandatory since the all-atom interaction
NNP is used in the subsequent step to perform the AHR
averages that yield the
reference energies
to parametrize
the AHR-averaged interaction NNP,
thus
allowing for an efficient evaluation of the AHR-averaged energies at
the
CC level of accuracy
but
without the need of performing hundreds of thousands of the
computationally
very demanding CC~calculations to obtain
the AHR-averaged interaction NNP; recall
that in order to compute each of the AHR reference energies,
$64 \times 32 = 2048$ interaction energies are needed, making the
direct
generation of the AHR-averaged NNP using explicit
\mbox{CCSD(T$^*$)-F12a/aVTZcp}
calculations
to evaluate each of them
computationally too expensive given that around 60~000
AHR energies are needed to
parametrize
the AHR-averaged interaction NNP
which includes here
the full flexibility of the solute.

\clearpage

\section{AHR-averaged Potential Energy Convergence}

In order compute the AHR-averaged interaction potential energies $V^{\rm{AHR}} (R, \theta, \phi) $, 
the adiabatic Schr\"{o}dinger
equation given by Eq.~(1) in the main text needs to be solved
for the lowest energy eigenvalue which provides $V^{\rm{AHR}}$
at the point $(R, \theta, \phi) $ that specifies the intermolecular configuration; 
see the middle panel of
Figure~1 in the main text for the definition of these spherical coordinates.
Solving this equation requires, on one side, the use of a linear combination of
spherical harmonics up to a maximum $l_\text{max}$ given by Eq.~(2) in the main text. 
On the other side, the evaluation of the potential energy contribution
requires a numerical integration using Gauss-Legendre and 
Gauss-Chebyshev
quadrature points to average over the $\theta^\prime$ and $\phi^\prime$~angles, 
respectively, according to Eq.~(3) in the main text.
The $l_\text{max}$ parameter as well as the number of Gauss-Legendre and 
Gauss-Chebyshev
quadrature points need to be determined beforehand, 
making sure that the obtained AHR-energy is converged
while not wasting computer time since the matrix elements
$ \langle Y_{l^\prime}^{m^\prime} | V(R,\theta,\phi,\theta^\prime,\phi^\prime) | Y_l^m \rangle $
need to be evaluated extremely often to implement the AHR-averaging
procedure based on the underlying all-atom interaction potential $V$ that
is represented in its turn by a corresponding all-atom interaction~NNP.

The convergence tests for these parameters have been performed using 
different configurations of both, \mbox{\wat{}$\cdots$\ph{}} and \mbox{\pwat{}$\cdots$\ph{}}
but for simplicity only the results obtained for
\wat{} is its minimum energy structure and the center of mass of \ph{}
at a distance $R=3.3$~\AA\ and angles $\theta=180^\circ$ and $\phi= 0^\circ$
according to the coordinates defined in the central panel of Figure~1 in the
main text are reported here.
In this section,
the AHR-averaged interaction energies have been
computed with the all-atom NNP and also with the explicit reference 
method,
where lots of \mbox{CCSD(T$^*$)-F12a/aVTZcp} single-point calculations 
for many different \h{} orientations are required to compute
the all-atom matrix elements
$ \langle Y_{l^\prime}^{m^\prime} | V(R,\theta,\phi,\theta^\prime,\phi^\prime) | Y_l^m \rangle $.
First, the number of Gauss-Legendre and
Gauss-Chebyshev 
quadrature points 
have been increased 
progressively 
from $2$ and $4$ to $128$ and $256$, respectively,
while fixing $l_\text{max}=8$. 
Here it is important to point out that the number of 
Gauss-Chebyshev
quadrature points 
is always 
twice the number of 
Gauss-Legendre quadrature points, 
as the $\phi^\prime$ interval is 
two times larger than
the $\theta^\prime$ interval, and this ensures
that the 
angular bin size remains the same for the two angles.

\begin{figure}[]
\centering
\includegraphics[width=\textwidth]{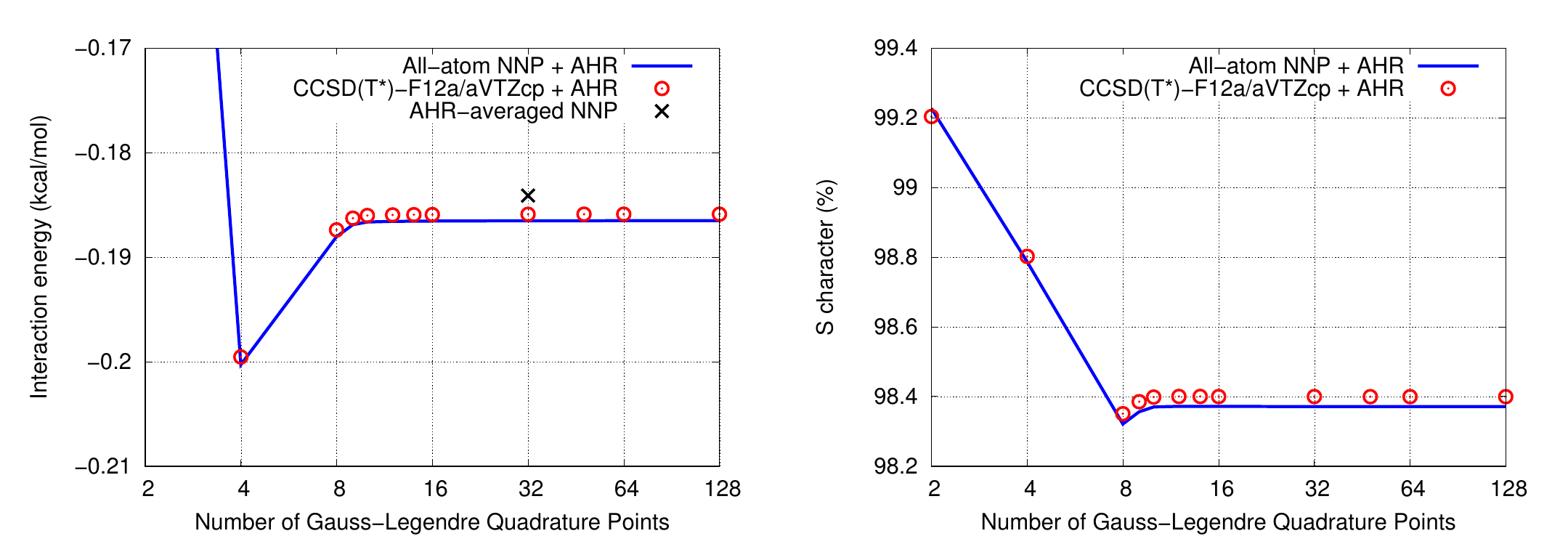}
\caption{
Convergence study of the AHR-averaged energy (left) and the 
spherical character in percent (right)
for \wat{}$\cdots$\ph{} in the AHR minimum structure, see text, 
as a function of the number of 
Gauss-Chebyshev 
and Gauss-Legendre quadrature points
used for the integration of the 
interaction potential 
energy matrix elements with respect to $\phi^\prime$
and $\theta^\prime$, respectively, 
whereas 
$l_{\text max}$ is set to~$8$.
The black cross marks the interaction energy as directly obtained from 
the AHR-averaged NNP at the accuracy level used to generate it. 
For simplicity, only the number of number of Gauss-Legendre quadrature points are
reported in the axis of the graph,
but
the number of 
Gauss-Chebyshev 
quadrature points is indirectly 
provided since is it twice the number of Gauss-Legendre quadrature points, see text.
Note that the blue line connecting the points of the all-atom NNP + AHR averaging
is a guide to the eye.
}
\label{fig: ahr minimum structure energy convergence}
\end{figure}

The convergence of the AHR-averaged energies with the number of 
quadrature points for the integration
of the potential matrix elements is shown in 
Figure~\ref{fig: ahr minimum structure energy convergence} (left), 
where it is possible to see that using $10$ to $12$ Gauss-Legendre
and $20$ to $24$ 
Gauss-Chebyshev 
quadrature points yields converged energies
on the relevant scale. 
Similarly, the percentage of spherical character of the structure is studied.
This percentage indicates how important the contribution of the spherical state
(i.e. $l=0$ and thus $m=0$) 
to the adiabatic hindered rotor average is. 
The smaller it is, the larger is the contribution of 
$l>0$ states 
and thus the more pronounced is the influence of anisotropy within the AHR-averaged energy.
In other words, 
the smaller the s-character the larger is the systematic 
error made when the simple spherical $l_{\rm max}=0$ average is used
to produce the effective interaction energy 
of the single-site \ph{} point particle from \mbox{all-atom \h{}.} 
The results are shown on Figure~\ref{fig: ahr minimum structure energy convergence} (right),
and also for this property only
$10-12$ Gauss-Legendre and $20-24$ 
Gauss-Chebyshev 
quadrature points
suffice to compute the converged value.

As already mentioned, the same convergence study has been performed for 
other \mbox{\wat{}$\cdots$\h{}} and \mbox{\pwat{}$\cdots$\h{}} configurations 
and it has been found that those structures with slightly repulsive
AHR-averaged energies require a larger number of quadrature points to 
achieve convergence, namely $32$ and $64$ quadrature points
for Gauss-Legendre and 
Gauss-Chebyshev, 
respectively.
Those structures also have a smaller s-character,
meaning that the 
$l>0$ states 
and thus 
anisotropies
contribute more to the AHR-averaged energies.
Indeed, they feature
a higher anisotropy on the all-atom potential energy surface
when performing rotations of the \h{} molecule around its center of mass,
which explains why more quadrature points are necessary in order to get
a converged AHR-averaged interaction potential energy.

\begin{figure}[]
\centering
\includegraphics[width=\textwidth]{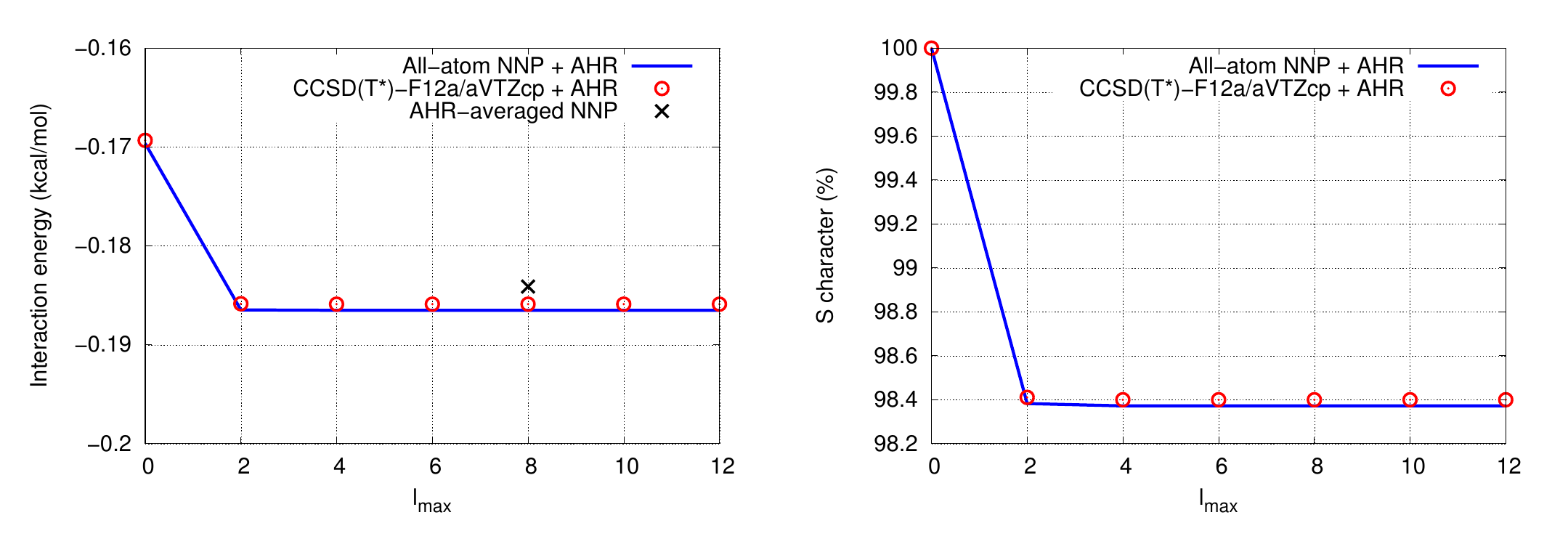}
\caption{
Convergence study of the AHR-averaged energy (left) and the
spherical character in percent (right)
for \mbox{\wat{}$\cdots$\ph{}} in the AHR minimum structure, see text,
as a function of $l_\text{max}$
used in the basis set expansion according to Eq.~(2) in the main text,
whereas the number of Gauss-Legendre and Gauss-Chebyshev quadrature
points are set to~$32$ and~$64$, respectively.
The black cross marks the interaction energy as directly obtained from 
the AHR-averaged NNP at the accuracy level used to generate it.           
Note that the blue line connecting the points of the all-atom NNP + AHR averaging
is a guide to the eye.
}
\label{fig: ahr minimum structure energy convergence l}
\end{figure}

After that, we carried out a second convergence study to determine $l_\text{max}$,
fixing the number of Gauss-Legendre quadrature points to $32$ and the 
Gauss-Chebyshev 
quadrature points to $64$, while now increasing $l_{\text max}$ 
from $0$ to $12$ in steps of $2$
(together with using all corresponding $m$~quantum numbers). 
Note that since we are considering \ph{} here, the AHR-average only 
includes even $l$ states. 
The results obtained for the energy and the spherical character are presented in 
Figure~\ref{fig: ahr minimum structure energy convergence l}. 
In this case, $l_\text{max}=2$ includes almost all the anisotropy of the structure,
but going up to $l_\text{max}=4$ is required to reach full convergence.
It needs to be taken into account, though, that this structure has an s-character
on the order of $98.4\%$ and, therefore, is not highly anisotropic.
Similarly to the number of quadrature points, the $l_\text{max}$ convergence has
also been tested with other structures, and those with a lower s-character require
up to $l_\text{max}=8$ to reach convergence.
Therefore, unless otherwise reported, all the AHR-averaged energies are computed
using $l_{\rm max}=8$
in conjunction with
$32$~Gauss-Legendre and 
$64$~Gauss-Chebyshev 
quadrature points.

Finally, Figures~\ref{fig: ahr minimum structure energy convergence} 
and~\ref{fig: ahr minimum structure energy convergence l}
not only report the AHR-averaged energy computed from the 
all-atom interaction NNP and the \mbox{CCSD(T$^*$)-F12a/aVTZcp} method,
but also the interaction energy 
that is directly
predicted by the AHR-averaged NNP
using the converged $l_{\text max}$ and quadrature point parameters,
see the black crosses. 
Therefore, this is a good estimate of the overall accuracy of the methodology
presented in this paper.
For this particular structure, the difference between the AHR-averaged NNP
interaction energy and the AHR-averaged energies computed from CC or from
the all-atom interaction NNP is of the order of 
$0.002$~kcal/mol.

\clearpage

\section{AHR-averaged \wat{}$\cdots${\textit{\MakeLowercase{p}}}H$_2$ Interaction NNP}

The next step in our data-driven procedure consists in generating the AHR-averaged 
interaction NNPs.
The accuracy of these AHR-averaged NNP has been tested by explicitly comparing 
the obtained AHR-averaged interaction energies from the AHR-averaged NNP 
to the interaction energies obtained 
while performing explicitly
the AHR~average 
using 
the all-atom NNP
along
different radial, angular and two-dimensional cuts of the respective potential energy surface.
It 
could not be
tested against the AHR-energies obtained with the 
electronic structure CC~method underlying the all-atom NNP due to 
the outrageous
computational effort
that would be involved 
to carry out these \mbox{CCSD(T$^*$)-F12a/aVTZcp} calculations 
as discussed in detail in the main text. 
Note that the AHR average of the all-atom NNP is performed
using the same $l_\text{max}$ and number of Gauss-Legendre and 
Gauss-Chebyshev
quadrature points as 
when generating
the AHR-averaged NNP
itself, 
namely 8, 32 and 64, respectively.

\begin{figure}[]
\centering
\includegraphics[width=\textwidth]{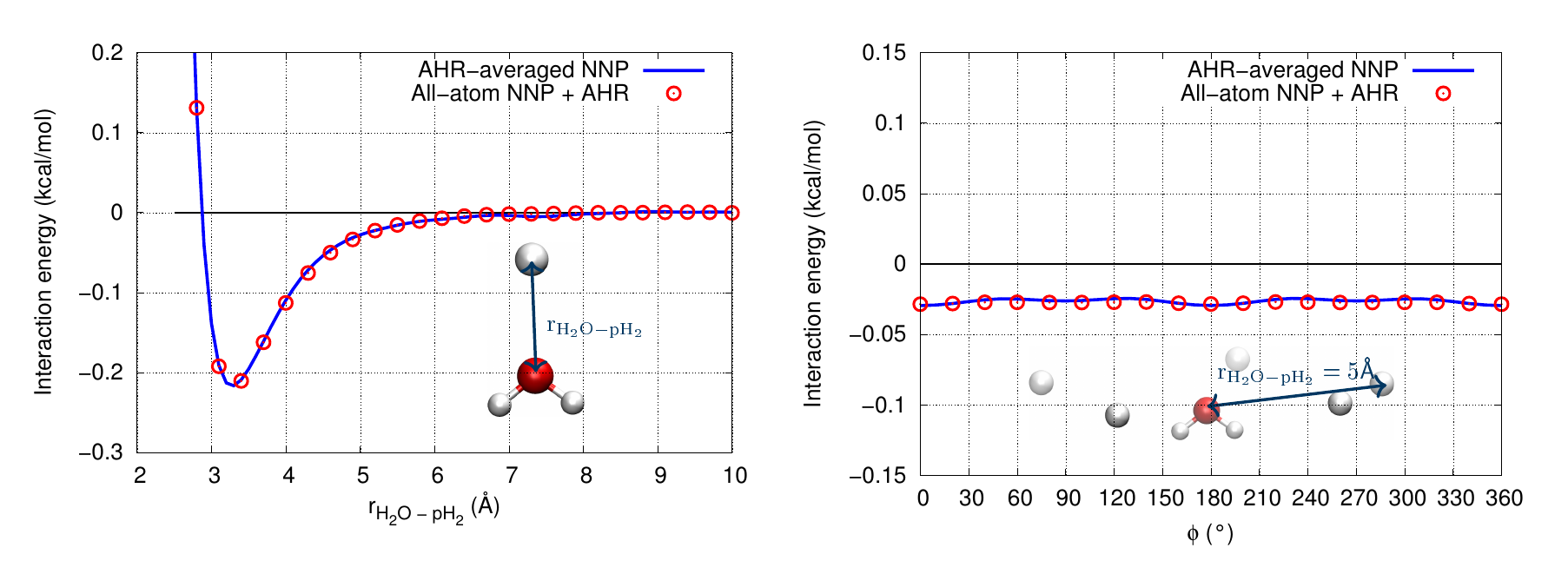}
\caption{
Comparison between the AHR-averaged interaction energies
obtained with the AHR-averaged NNP and
performing the AHR average with the all-atom NNP
for a radial (left) and an angular (right) cut of the \mbox{\wat{}$\cdots$\ph{}} system.
The configurations used for these cuts are provided in the insets. }
\label{fig: ahr-nnp radial angular scan}
\end{figure}

\begin{figure}[]
\centering
\includegraphics[width=\textwidth]{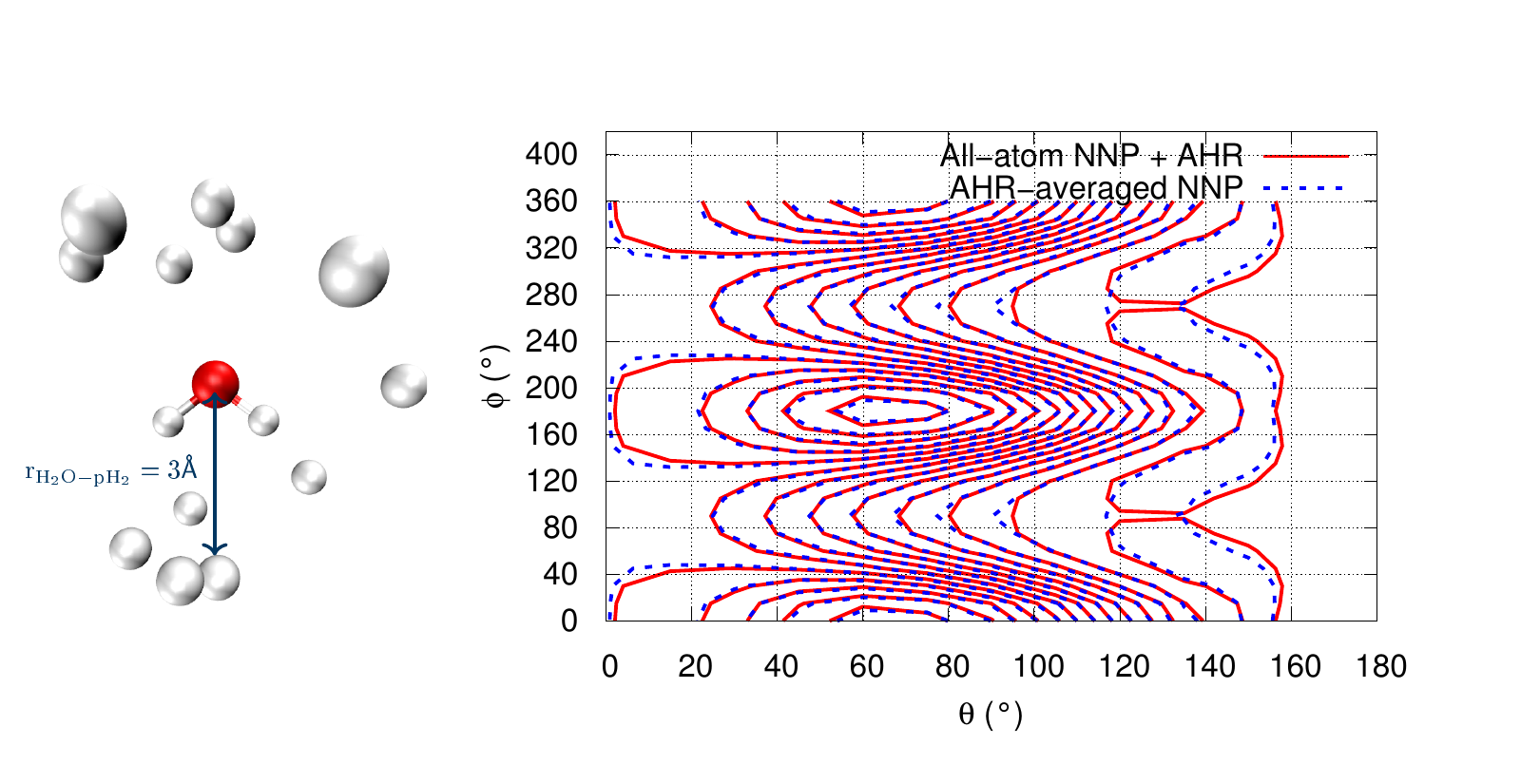}
\caption{
Comparison of the contour plots obtained with the AHR-averaged interaction
energies obtained with the AHR-averaged NNP and performing the AHR average using the
all-atom NNP for a two-dimensional angular scan of
the \mbox{\wat{}$\cdots$\ph{}} system.
The configuration used for this scan is provided in the left panel
and
energy scale corresponding to this contour plot
ranges from
$-$0.25 to 0.05~kcal/mol. }
\label{fig: ahr-nnp 2d scan}
\end{figure}

The result for the radial scan along the $C_{2v}$ axis of the water molecule
is shown in \mbox{Figure~\ref{fig: ahr-nnp radial angular scan} (left),}
and an angular scan 
(at the fixed intermolecular distance as indicated in the inset)
is shown in the same figure 
in the right panel. 
In both cases the agreement between the AHR-interaction energy computed from
the all-atom NNP 
after explicit AHR averaging 
and that predicted with the AHR-averaged NNP is almost perfect
in particular considering the energy scales. 
A similar but now two-dimensional 
angular
cut along the potential energy surface is shown in 
Figure~\ref{fig: ahr-nnp 2d scan}, where a very good accuracy can again be observed.

\begin{figure}[]
\centering
\includegraphics[width=0.9\textwidth]{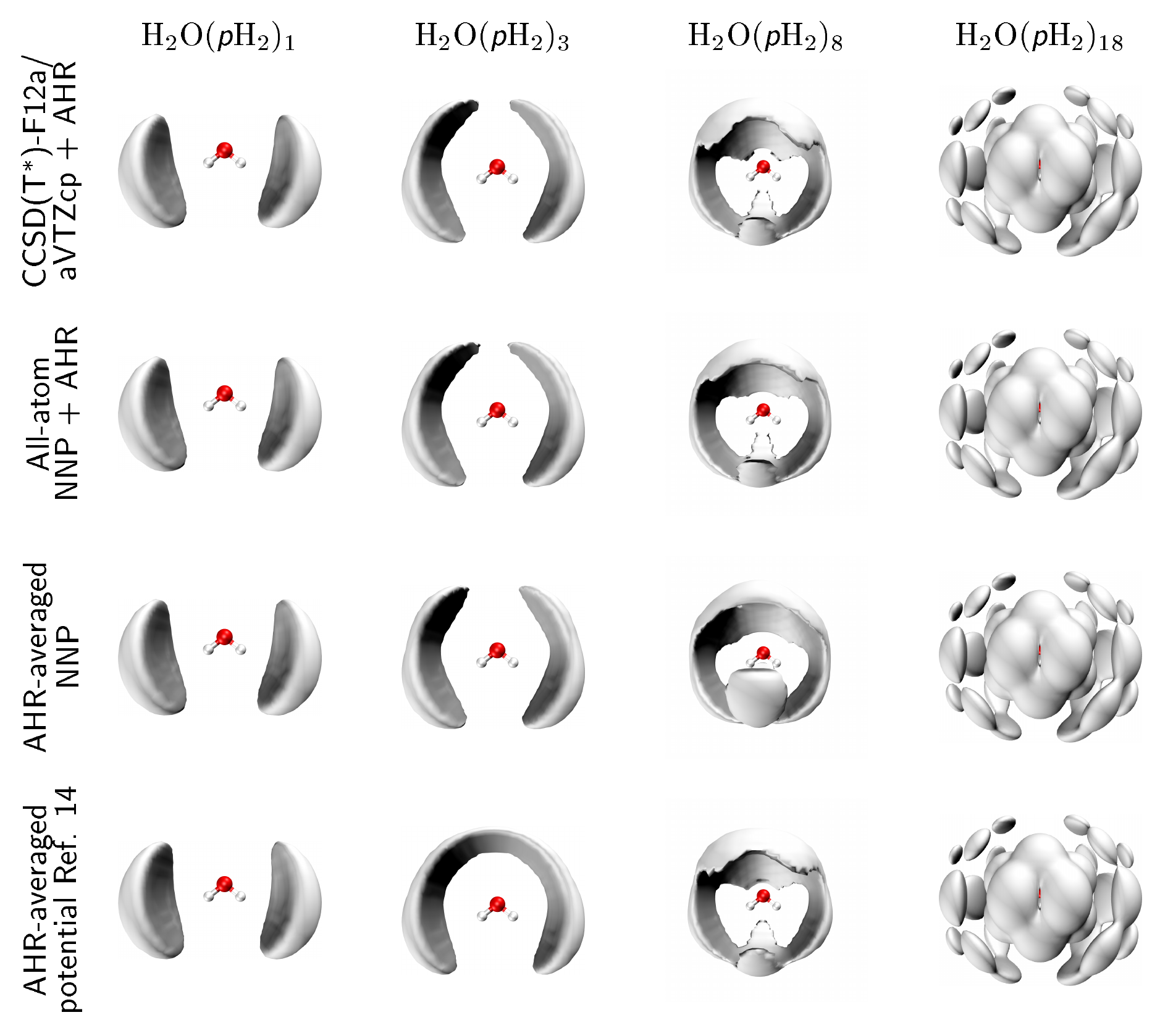}
\caption{
Comparison of the 
\ph{} SDFs around frozen \wat{} on a grid, see text, 
obtained with different methods to compute the interactions
for four
cluster sizes $N$.
From top to bottom:
(i) \mbox{CCSD(T$^*$)-F12a/aVTZcp} + AHR-averaging,
(ii) all-atom \mbox{\wat{}$\cdots$\ph{}} NNP + AHR averaging,
(iii) AHR-averaged \mbox{\wat{}$\cdots$\ph{}} NNP,
and 
(iv) AHR-averaged potential from Ref.~\citenum{
zeng2011adiabatic}. 
From left to right: $N=1$, 3, 8 and 18 \ph{} molecules 
solvating the minimum energy structure of the water molecule. 
The isovalues are 0.0009, 0.0038, 0.0075, and 0.0015~1/bohr$^3$ for $N=1$, 3, 8 and 18.
}
\label{fig: ahr-nnp h2o sdf}
\end{figure}

Last but not least, we report the computed spatial distribution functions (SDFs)
from 
bosonic
path integral simulations 
at a temperature of 1~K
for \mbox{\wat{}(\ph{})$_N$} complexes with 
$N=1,3,8$ and $18$ \ph{} in Figure~\ref{fig: ahr-nnp h2o sdf}.
Similarly to the main text, the SDFs are computed on a grid for a \wat{}
molecule fixed at its minimum energy structure using 
(i)
the reference CC method and performing the AHR-average on the obtained energies,
(ii)
performing the AHR-average while using the all-atom NNP interaction energies,
(iii)
directly using
the AHR-averaged NNP, 
and 
(iv)
using the published AHR-averaged potential from Ref.~\citenum{
zeng2011adiabatic}.
Overall, there are no significant deviations between the SDFs
obtained with the four different methods, in agreement with the
rest of the cluster sizes presented in the main text.

\clearpage

\clearpage
\section{All-atom \pwat{}$\cdots$\h{} Interaction NNP}

The methodology presented in this paper 
as detailed herein so far for \wat{}
has also been applied to generate the
all-atom and AHR-averaged interaction NNPs to describe the 
\mbox{\pwat{}$\cdots$\ph{}}
interaction.
Therefore, in this and the following sections, we will 
validate
the accuracy of the obtained interaction NNPs
for the hydronium cation.

In the first step, the all-atom NNP is generated where 
the 
rotational degrees of freedom of \h{} are included explicitly.
This all-atom \mbox{\pwat{}$\cdots$\h{}} interaction NNP has been carefully tested
by explicitly comparing the energies predicted by the all-atom NNP to the 
reference \mbox{CCSD(T*)-F12a/aVTZcp} calculations for some cuts along the potential energy
surface.
Such an approach is limited to small sections of the configuration space and
it is therefore complementary to the more meaningful and complex testing
shown in the main text using 
SDFs 
computed from path integral simulations.
We want to highlight that the main focus of this testing is on the
rotation of the \h{} molecule around 
its center of mass given a configuration of
the \pwat{} solute, as this 
is very important
for the AHR~averaging procedure and will 
ultimately limit
how accurate this averaging 
can be.

\begin{figure}[]
\centering
\includegraphics[width=\textwidth]{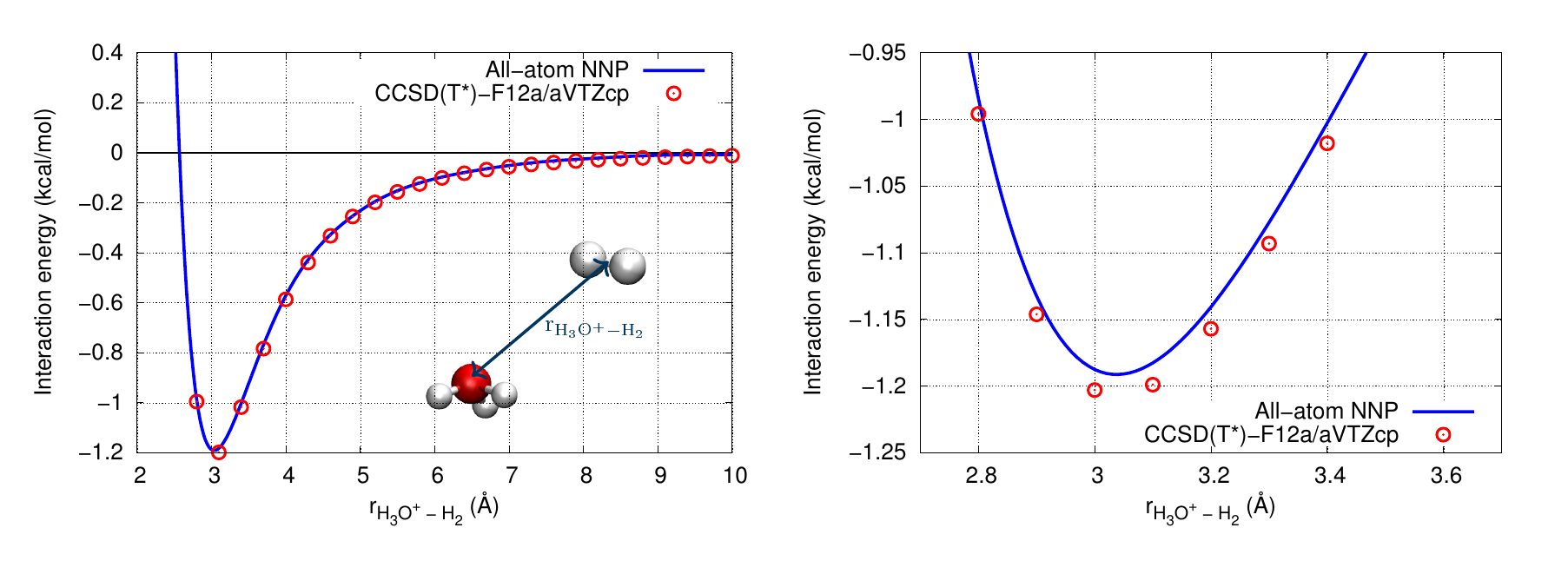}
\caption{
Comparison 
between the
interaction energies computed with the reference
\mbox{CCSD(T$^*$)-F12a/aVTZcp} method and the all-atom NNP for a radial scan
of \pwat{}-\h{} (see inset).
The right panel 
magnifies
the region around the
minimum of the interaction energy 
profile.
}
\label{fig: radial full dimensional h3o}
\end{figure}

\begin{figure}[]
\centering
\includegraphics[width=\textwidth]{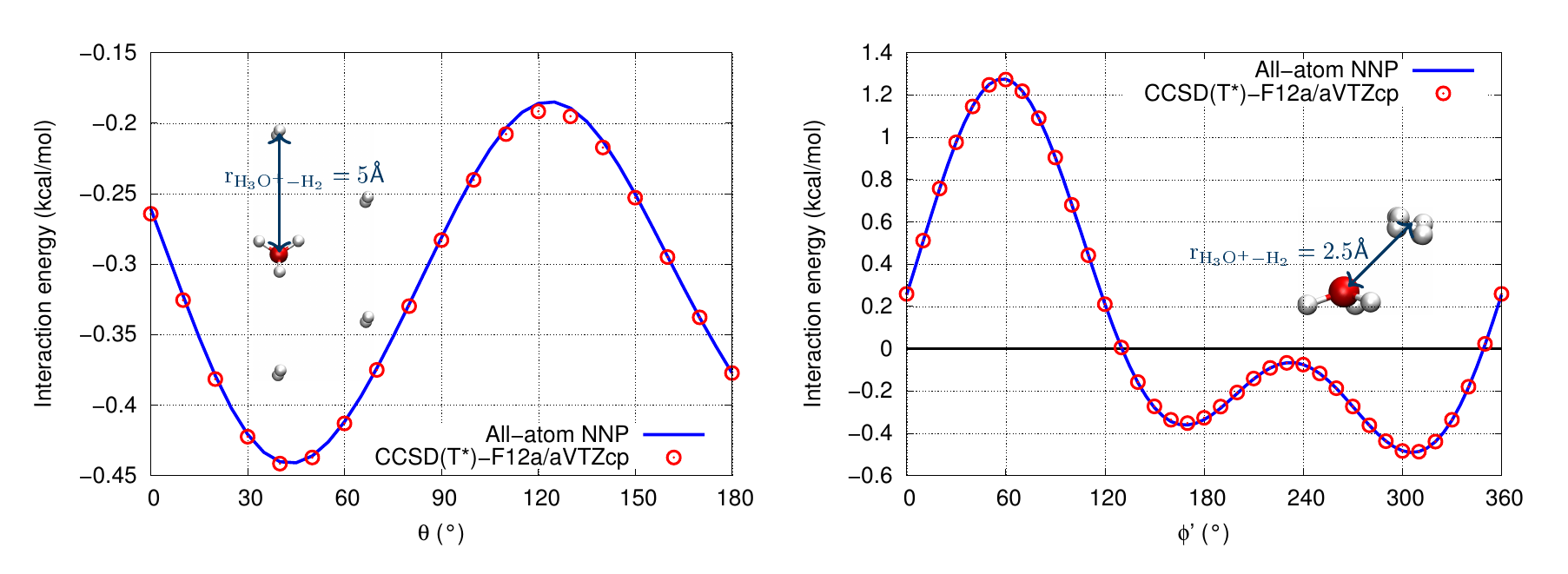}
\caption{
Comparison between the interaction energies computed with the reference
\mbox{CCSD(T$^*$)-F12a/aVTZcp} method and the all-atom NNP for 
two different conformations of the \mbox{\pwat{}$\cdots$\h{}} system (left and right)
as a function of 
the rotation of the center of mass of the \h{} molecule around the
\pwat{} molecule along the $\theta$ angle (left) and
the rotation of the \h{} molecule around its center of mass 
along the $\phi^\prime$ angle (right).
}
\label{fig: angular full dimensional h3o}
\end{figure}

The results for a radial scan are given in Figure~\ref{fig: radial full dimensional h3o}.
On 
the relevant scale of the interaction potential energy surface (left panel)
it is found that essentially 
no differences between the all-atom NNP and the 
CC~interaction energies can be seen.
However, slight differences between the CC and the all-atom NNP
are observed in the zoom-in that is depicted in the right panel.
These small deviations are 
about $0.015$~kcal/mol 
and,
thus, are considered negligible on the relevant energy landscape. 
A similar comparison has also been carried out for scans along the rotational
degrees of freedom: Figure~\ref{fig: angular full dimensional h3o} shows
the results for two of them, one for the 
rotation of the center of mass of the \h{} molecule
around \pwat{} (left panel) and another one for the rotation of the \h{} molecule
around its center of mass (right panel);
the same coordinate system as introduced in the middle panel of Figure~1 in the main text is used. 
Both of them show a convincing agreement between the interaction energies predicted
with the all-atom NNP and those computed explicitly with the 
\mbox{CCSD(T*)-F12a/aVTZcp}
reference
method.

\begin{figure}[]
\centering
\includegraphics[width=\textwidth]{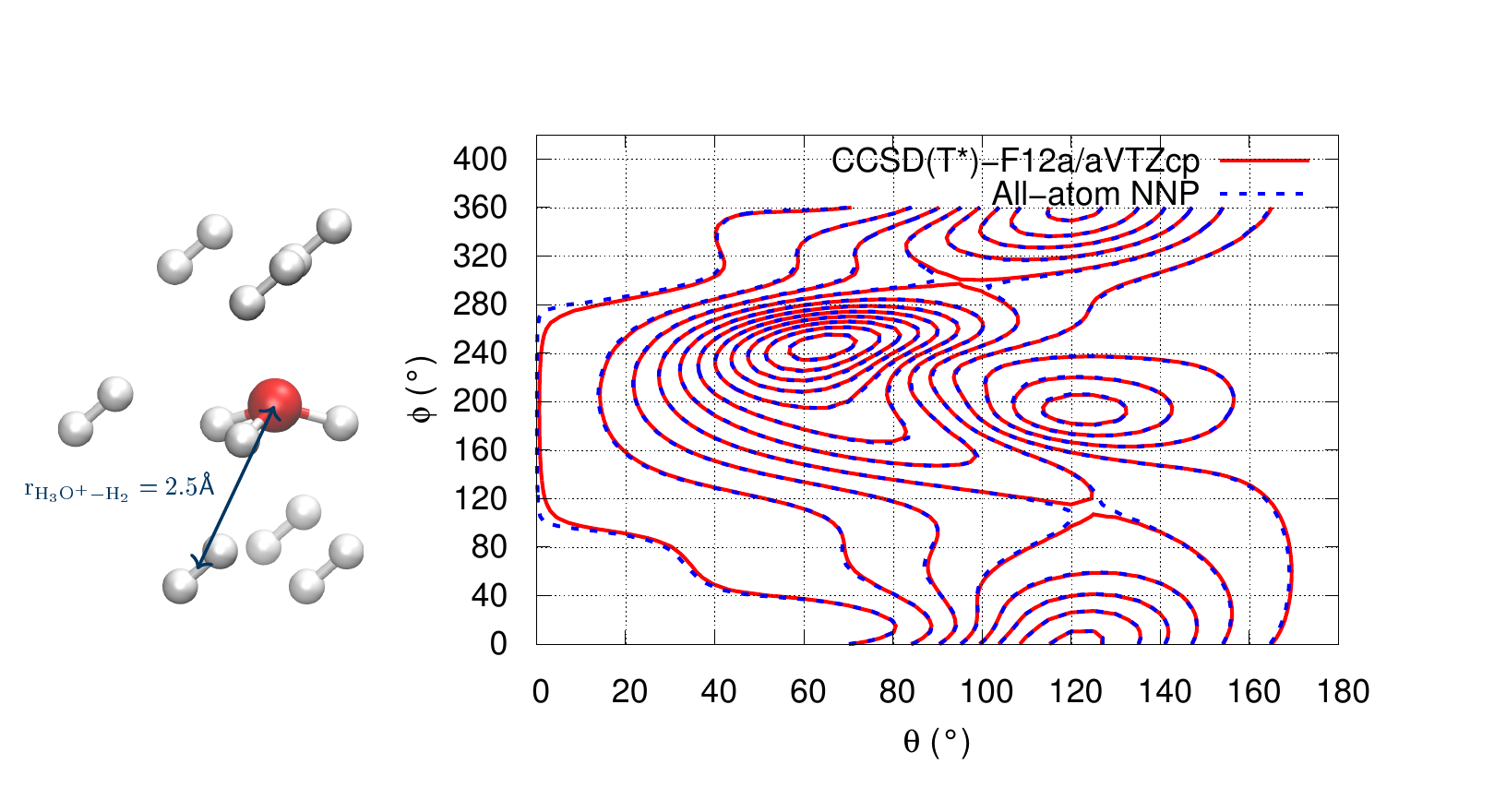}
\caption{
Comparison of the contour plots obtained with the all-atom NNP  
interaction energies
and the \mbox{CCSD(T$^*$)-F12a/aVTZcp} interaction energies
for a 2-dimensional scan over the orientation of the \h{} molecule around
the \pwat{} molecule
at the global minimum energy structure,
where
the orientation of \h{} 
with respect to its center of mass
is kept fixed.
The energy scale corresponding to this contour plot is 
$-$10 to 4~kcal/mol.
}
\label{fig: 2d full dimensional h3o}
\end{figure}

\begin{figure}[]
\centering
\includegraphics[width=\textwidth]{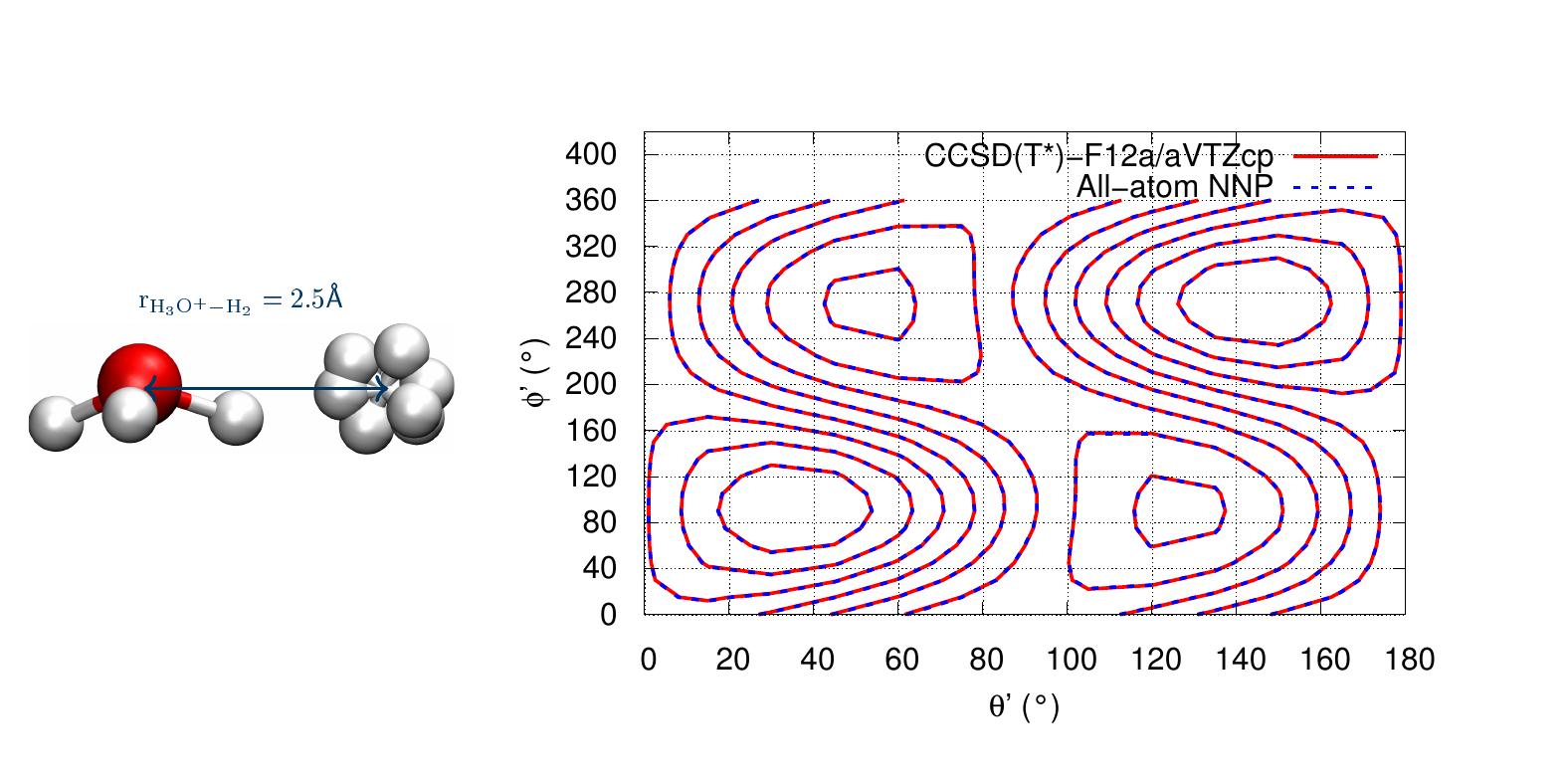}
\caption{
Comparison of the contour plots obtained with the all-atom NNP 
interaction energies
and the \mbox{CCSD(T$^*$)-F12a/aVTZcp} interaction energies
for a 2-dimensional scan over the
orientation of the \h{} molecule around its center of mass, 
with the \pwat{} molecule at the global minimum energy structure
keeping the \h{} center of mass fixed.
The energy scale corresponding to this contour plot is 
$-$3 to 1.5~kcal/mol.
}
\label{fig: 2d full dimensional h3o 2}
\end{figure}

Finally, also two-dimensional cuts of the potential energy surface have been analyzed
for the \pwat{} case.
Figure~\ref{fig: 2d full dimensional h3o} illustrates the quality of the fit
when changing the orientation of the \h{} molecule around the \pwat{} molecule
but keeping fixed the orientation of the \h{} around its center of mass,
whereas Figure~\ref{fig: 2d full dimensional h3o 2} shows the opposite,
i.e. the \h{} molecule rotates around its center of mass and this one has a
fixed orientation with respect to the \pwat{} molecule.
But in both cases, the agreement between the all-atom interaction NNP
and the CC~reference method is almost perfect.

\clearpage
\section{AHR-averaged \pwat{}$\cdots${\textit{\MakeLowercase{p}}}H$_2$ Interaction NNP}
As for water, 
the all-atom NNP generated to describe the \mbox{\pwat{}$\cdots$\h{}} 
interaction can be used to obtain the AHR-averaged NNP.
In order to test the accuracy of the AHR-averaged network, an explicit comparison
between the interaction energies predicted by the AHR-averaged NNP and 
the AHR-averaged energies computed from the all-atom NNP is performed
for the hydronium cation.
Again,
the AHR interaction energies are computed using the same number of 
quadrature points and $l_\text{max}$ that have been used to generate the AHR-averaged NNP, 
so 32~Gauss-Legendre and 64~Gauss-Chebyshev quadrature points 
together with $l_\text{max}=8$.

\begin{figure}[H]
\centering
\includegraphics[width=\textwidth]{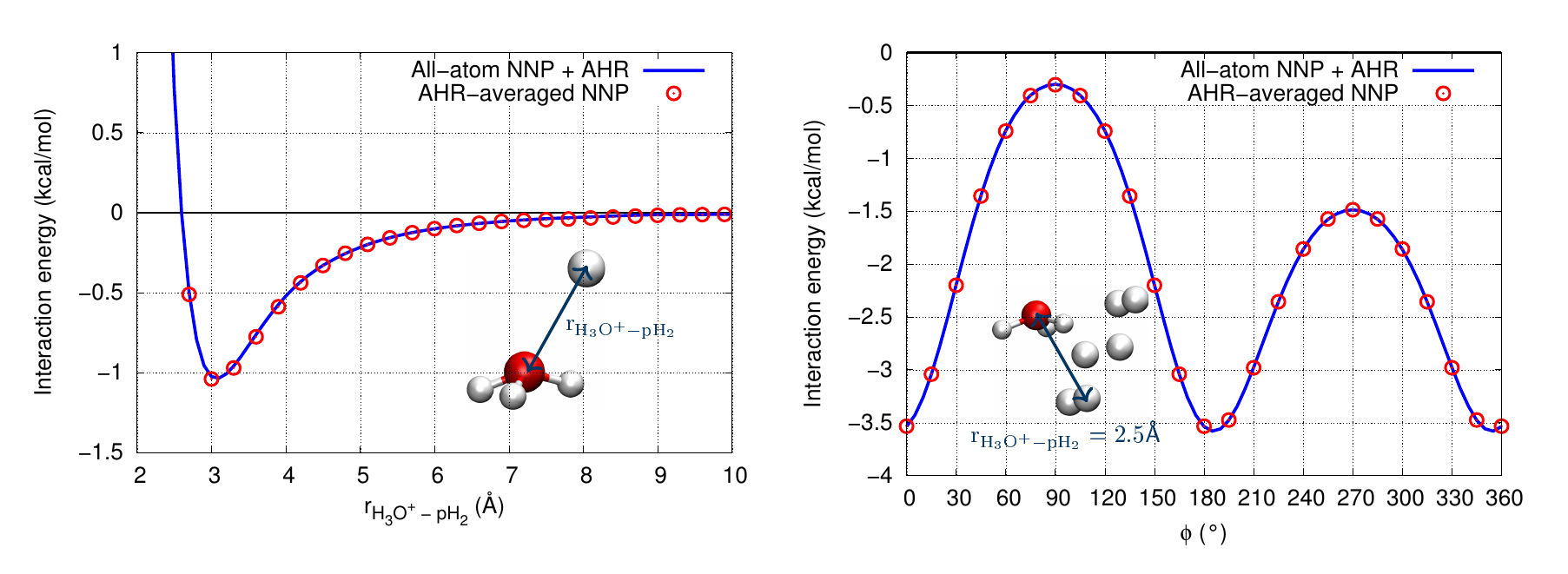}
\caption{
Comparison between the AHR-averaged interaction energies 
obtained
with the AHR-averaged NNP and 
performing the AHR-average 
with
the all-atom NNP for a radial (left)
and an angular scan (right) of the \mbox{\pwat{}$\cdots$\ph{}} system.
The configurations used for these scans are provided in the insets.
}
\label{fig: angular ahr nnp h3o}
\end{figure}

\begin{figure}[H]
\centering
\includegraphics[width=\textwidth]{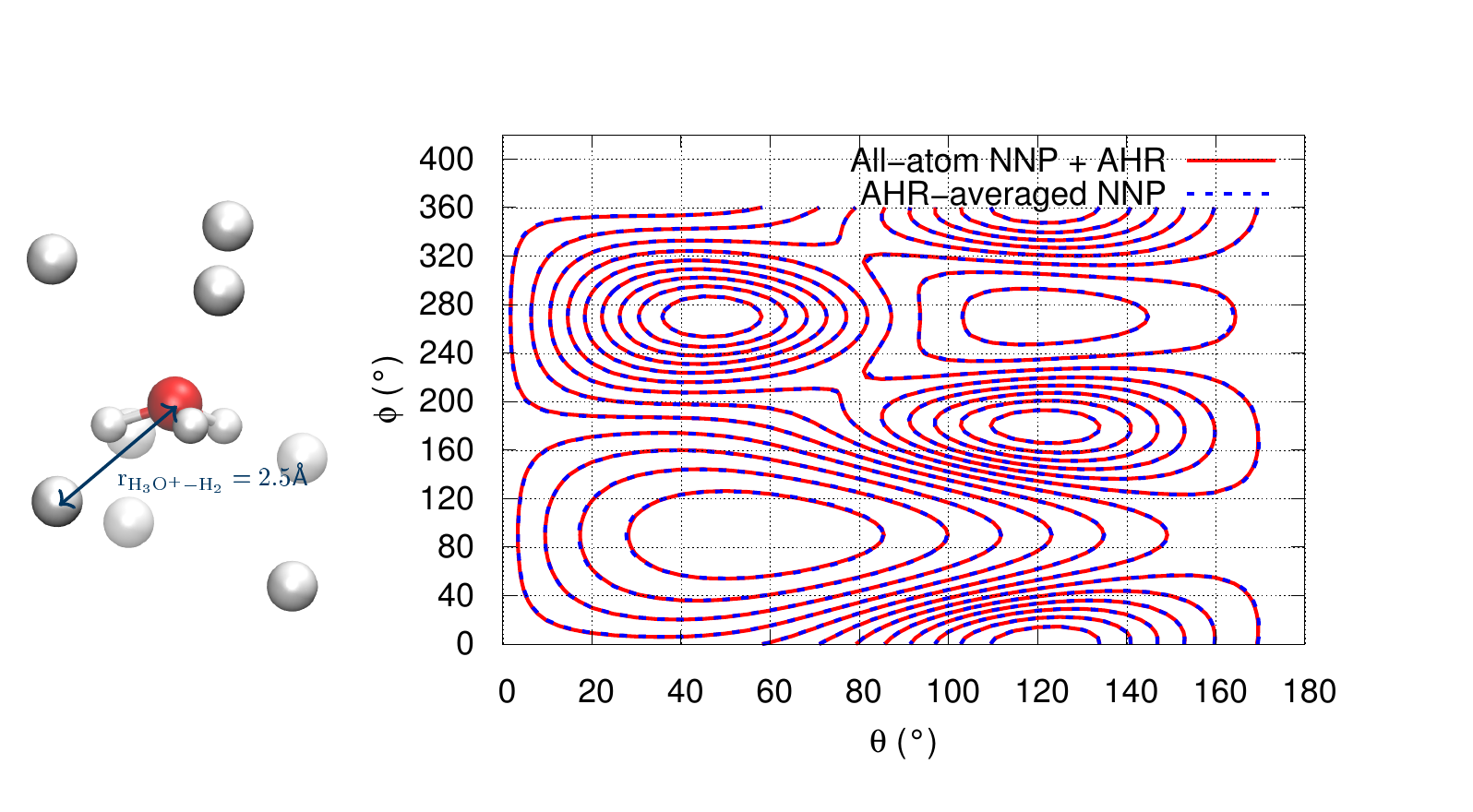}
\caption{
Comparison of the contour plots obtained with the AHR-averaged
interaction energies obtained with the AHR-averaged NNP and performing
the AHR average 
using the all-atom NNP 
for a 2-dimensional angular scan of the \mbox{\pwat{}$\cdots$\ph{}} system.
The configuration used for this scan is provided in the left panel and
the energy scale corresponding to this contour plot ranges from $-$4 to 0~kcal/mol. 
}
\label{fig: 2d ahr nnp h3o}
\end{figure}

The results of these tests of the AHR-averaged NNP are shown in 
Figure~\ref{fig: angular ahr nnp h3o} 
as radial (left panel) and angular (right) 
scans and 
in Figure~\ref{fig: 2d ahr nnp h3o}
as a two-dimensional scan in the two angular degrees of freedom
using the coordinate system from the middle panel of Figure~1 in the main text
and the configurations depicted in these figures. 
In all three 
representative
cases,
the agreement between the averaged energies
predicted by the AHR-averaged NNP and those computed by performing the
explicit 
AHR-average with the all-atom NNP is almost exact.

\begin{figure}[]
\centering
\includegraphics[width=1.0\textwidth]{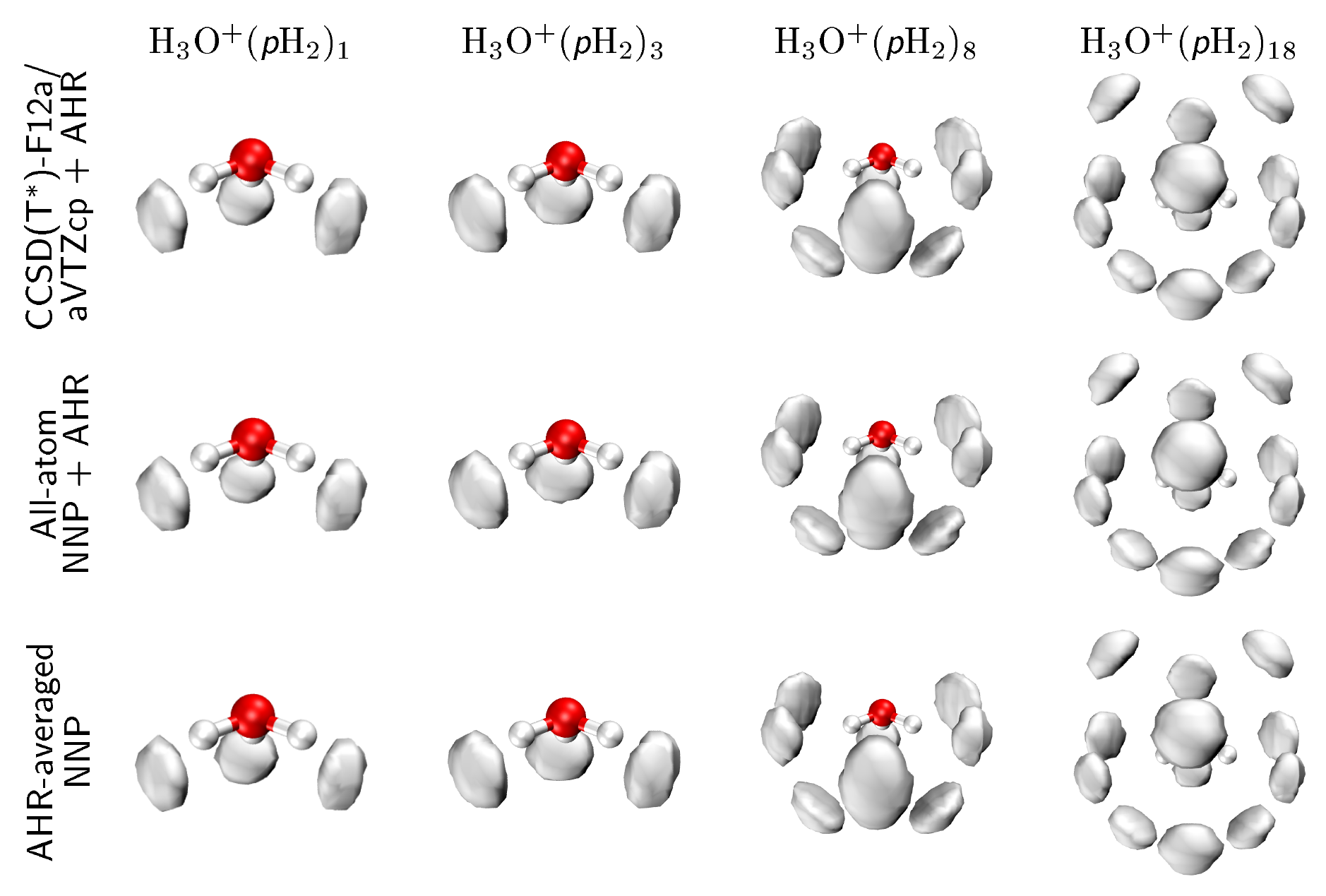}
\caption{
Comparison of the \ph{} SDFs around frozen \pwat{} on a grid, see text,
obtained with different methods to compute the interactions for four cluster sizes $N$.
From top to bottom:
(i) \mbox{CCSD(T*)-F12a/aVTZcp} + AHR averaging,
(ii) all-atom \mbox{\pwat$\cdots$\h{}} NNP + AHR averaging,
and
(iii) AHR-averaged \mbox{\pwat$\cdots$\ph{}} NNP.
From left to right:
$N=1$, 3, 8 and 18 \ph{} molecules solvating the minimum energy structure
of the hydronium cation.
The isovalues are 0.0250 for $N=1$, 3 and 8 and 0.0460 for $N=18$ \ph{}.
}
\label{fig: ahr-nnp h3o sdf}
\end{figure}

Also for \pwat{} we report in Figure~\ref{fig: ahr-nnp h3o sdf}
the obtained SDFs
for the \pwat{}(\ph{})$_N$ clusters
for $N=1,\ 3,\ 8$ and $18$ \ph{} from left to right. 
The SDFs are computed on a grid keeping the \pwat{} molecule fixed 
according to its minimum energy structure 
using 
(i)
the CC reference method and performing the AHR-average on the obtained energies
(top row), 
(ii) performing the AHR-average using the all-atom NNP interaction energies
(middle row), 
and 
(iii) using directly the final AHR-averaged interaction NNP
(bottom row). 
Overall, there are no significant deviations between the SDFs
of \ph{} around \pwat{} 
obtained with the three different methods in agreement with the
other the cluster sizes presented on the main text.

\clearpage

\clearpage